\documentclass[9pt,twocolumn,twoside]{osajnl}
\usepackage{subfigure}

\journal{josab} % Choose journal (ao, aop, josaa, josab, ol, pr)

\title{Inverse design  of a 1D dielectric metasurface by topology optimization; fluctuations-trend analysis assisted by a diamond-square algorithm}

\author[1,*]{K. Edee}
\author[1,2]{M. Ben Rhouma}
\author[3]{J.-A. Fan}
\author[2]{M. Antezza}
\author[4]{N. Gippius}
\author[3]{E. Wang }
\author[1]{J.-P.  Plumey}
\author[1]{G. Granet}
\author[2]{B. Guizal}

\affil[1]{Universit\'{e} Clermont Auvergne, Institut Pascal, BP 10448, F-63000 Clermont-Ferrand, France, CNRS, UMR 6602, Institut Pascal, F-63177 Aubi\`{e}re, France}
\affil[2]{Laboratoire Charles Coulomb (L2C), UMR 5221 CNRS-Universit\'{e} de Montpellier, F-34095 Montpellier, France}
\affil[3]{Department of Electrical Engineering, Stanford University, Stanford, California 94305, United States}
\affil[4]{Skolkovo Institute of Science and Technology, Nobel Street 3, Moscow 143025, Russia}

\affil[*]{Corresponding author: kofi.edee@uca.fr}

\begin{abstract}
 We present a topology optimization (TO) method for a 1D dielectric metasurface, coupling the classical trend-fluctuations analysis (FTA) and the diamond-square- algorithm (DSA). 
In the classical FTA, a couple of device distributions  termed  Fluctuation or mother and Trends or father, with specific spectra is initially generated. The spectral properties of the trend function, allow to target efficiently the basin of optimal solutions.
For optimizing a 1D metasurface to deflect a normally incident plane wave into a given deflecting angle,  a cosine-like function has been identified to be an optimal father profile allowing to efficiently target  a basin of local  minima. However there is no efficient method to predict the father profile number of oscillations  that effectively allows to avoid undesirable local optima. It would be natural to suggest a randomization of the variable which controls the number of oscillations of the father function. 
However, one of the main drawbacks of the randomness searching process is that, combined with a gradient method, the  algorithm can  target, undesirable local minima.  
The method proposed in this paper improves the possibility of the classical FTA to  avoid  the trapping of undesirable local optimal solutions. This is accomplished  by extending the initial candidate family to higher quality offspring that are generated thanks to a diamond-square-algorithm (DSA). Doing so, ensures that the main features of the best trends are stored in the genes of all Offspring structures.   
\end{abstract}

\setboolean{displaycopyright}{true}

\begin{document}
\maketitle

%%%%%%%%%%%%%%%%%%%%%%%%%%  body  %%%%%%%%%%%%%%%%%%%%%%%%%%
\section{Introduction and statement}
%%%%%%%%%%%%%%%%%%%%%%%%%%  body  %%%%%%%%%%%%%%%%%%%%%%%%%%
%Optimizers based on inverse design with a Topology optimization (TO) process, have newly received significant attention in photonics  \cite{Sell1, Yang1, Sell2, Yang, Phan, EWang}.  This optimization process can be roughly viewed as a concatenations of: initialization, evaluation and updating process. During the initialization, initial trivial-shape or random-shape   candidates are generated as initial populations of the optimizer. In the evaluating stage, the fitness of these initial populations is evaluated through  the figure of merit (FOM) function,  which can be  efficiently computed through the so-call adjoint method. This stage can also be viewed as an exploration stage.   Finally a new population/design variable is updated during the updating stage.
%randomly performs the exploration process and exploitation process.
%In some  works \cite{Sell1, Yang1, Sell2, Yang, Phan, EWang}, previously highlighted that algorithms based on initial random candidates yield  higher-quality solutions than those based on initial geometric layouts with canonical shape.\\
Inverse design algorithms based on the topology optimization (TO) process have received significant attention in photonic device design \cite{Lu, Lalau, Hughes,Molesky,Frandsen, Borel, Piggot, Xiao, Lin, Sell1, Yang1, Sell2, Yang, Phan, EWang}. The optimization process consists of three mains parts: initialization, evaluation, and updating. During initialization, devices are set to have trivial or random dielectric distributions.
These devices are then iteratively evaluated using the adjoint variables method, which yields gradients that specify how the device can improve a given figure of merit (FOM), and updated using gradient descent.  TO has been successfully used to produce a broad range of freeform photonic devices operating in free space and on-chip, including metasurfaces, metagratings, wavelength routers, optical isolators, and photonic crystals \cite{Lu, Lalau, Hughes,Molesky,Frandsen, Borel, Piggot, Xiao, Lin, Sell1, Yang1, Sell2, Yang, Phan, EWang}. As a gradient-based optimizer, TO is a local optimizer that is only capable of searching a limited region of the total design space, which is vast and non-convex. There are  different strategies allowing to  more thoroughly search this space. Among them, a  global optimum-inspired scenarios \cite{Jiang} and a well-defined initial condition strategy. 
In the second scenario, multiple optimizations are typically performed with differing initial dielectric distributions.  An ensemble of locally optimized devices with a range of capabilities is produced, and the highest performing device is selected as the final device pattern.  While this method works, it is computationally expensive and still limited to a set of local optimized devices, with limited search capabilities of the design space.  
However, an appropriate prediction of initial geometry features,  can eliminate the drawbacks by efficiently targeting a basin of optimal local minimum and significantly reducing the computation time.\\
The Fluctuation and Trends analysis (FTA), which combines the use of local gradients with an appropriate prediction of initial geometry features has been recently proposed \cite{{Kofi_TO}} and it appears as a promising route avoiding a blind scan of the design space. 
In this optimization scheme, two basic patterns with desired statistic properties are generated as initial  devices geometry. The mother profile (or fluctuating function) holds the random fluctuations while the father provides the initial trend. The  spectrum of the father distribution provides the main features of desired final optimized binarized device.
In the case of a 1D functional metasurface,  a cosine-like function has been identified to be an optimal initial father profile yielding a  desired final full-binarized structure. However there is no efficient method to predict the father profile number of oscillations  that effectively allows to avoid undesirable local optima. Randomization of the variable controlling the father  function shape is then required. But this stochastic process would be interesting only if it doesn't lead to a prohibitive cost in terms of computing time.\\ 
In this study, we explore a strategy termed  Fluctuation and Trends Analysis-Diamond Square Algorithm (FTA-DSA) , which combines the use of local gradients with an evolutionary-kind optimization. The proposed hybrid method is a promising route to expand TO to a global search of the design space.
The augmentation of FTA with the DSA \cite{Fournier,Miller, Lewis} allows to more systematically perform global optimization of photonic devices, by generating better initial candidates,  while preserving the advantages introduced by the use of primary initial sinusoidal father profile function.  
The DSA has been firstly proposed by  Fournier, Fussell and Carpenter \cite{Fournier} for the  generation of two-dimensional landscapes or procedural texture  with a very realistic-looking. It is commonly  used in computer graphics software to model a surface or volume distribution of natural elements. Generally, the natural appearance \textsl{i.e.} the numerical representation of the randomness of these natural distribution  is obtained by systematically performing interpolations of  spatial values of some given nodes augmented with a random fractal noise function. This procedure provides a distribution of variables presenting both a random rendered and a self-similarity properties. %provides landscapes or procedural texture  with a very realistic-looking. 
The FTA assisted by the DSA  works by taking multiple FTA-optimized devices, obtained after a very low number of iterations and systematically performing interpolations of the spatial characteristics of the best candidates of these devices to search within this design space region.  We apply this algorithm to the optimization of one-dimensional metagratings, which consist of polysilicon nanoridges and that diffract an  incident radiation into a given  diffraction order.  Metagratings are a good model system because they have a well-defined FOM, can be simulated accurately and with high speeds using solvers based on modal methods  \cite{Knop,Granet_Guizal,Lalanne,Li,Granet_ASR,Kofi_gegenbauer1,Kofi_gegenbauer2,Kofi_gegenbauer3,Kofi_2D, Kofi_elsevier}, and can be benchmarked with prior results.

%%%%%%%%%%%%%%%%%%%%%%%%%%  body  %%%%%%%%%%%%%%%%%%%%%%%%%%
\section{Methods}
%%%%%%%%%%%%%%%%%%%%%%%%%%  body  %%%%%%%%%%%%%%%%%%%%%%%%%%
%%%%%%%%%%%%%%%%%%%%%%%%%%  body  %%%%%%%%%%%%%%%%%%%%%%%%%%
\subsection{Basic concepts of the topology optimization by FTA}
\begin{figure}[htb!]
 \centering
         {\includegraphics[width=.5\textwidth]{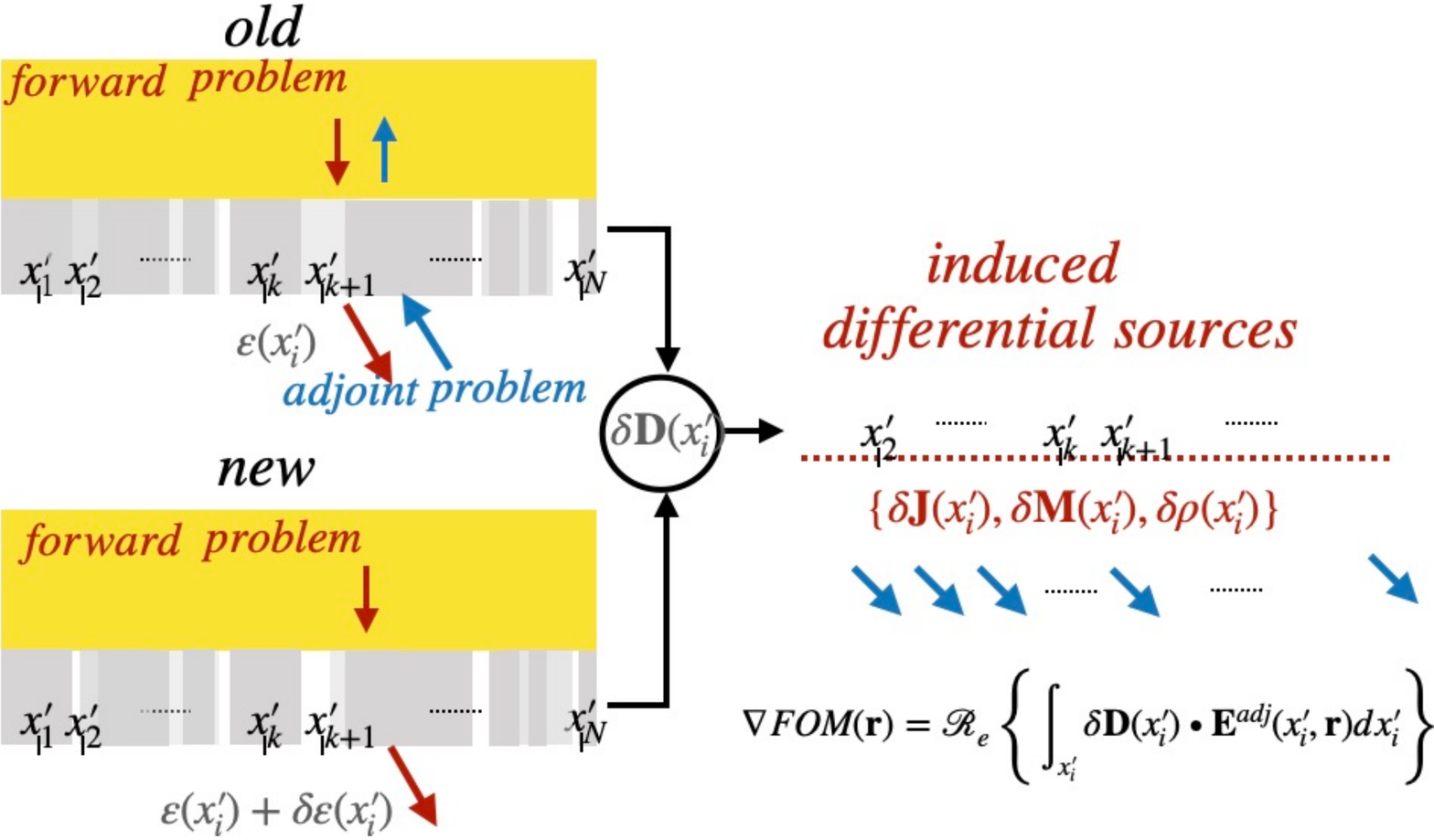}}
        %\caption{\label{deflection_GG_high} toto}
\caption{\label{sketch_ADM}
Flowchart of the adjoint method used to compute the gradient of the figure of merit.}
\end{figure}

%%%%%%%%%%%%%%%%%%%%%%%%%%  body  %%%%%%%%%%%%%%%%%%%%%%%%%%
The FTA schematized in fig.\ref{sketch_ADM}  is an iterative  adjoint-based topology optimization method generally coupled with a gradient algorithm and used in the design of photonic devices. 
The algorithm starts with two initial continuous permittivity profiles, termed mother profile and father profile, possessing particular spectra that prevent the algorithm, as much as possible, being trapped in local minima.
%%%%%%%%%%%%%%%%%%%%%%%%%%
%%%%%%%%%%%%%%%%%%%%%%%%%%
At iteration $t$, a {\it forward} and an {\it adjoint} (or reciprocal) computations (cf. fig.\ref{sketch_ADM}) are performed yielding two kinds of electromagnetic fields, termed {\it direct field} and {\it adjoint field} respectively. These fields are then used to compute, (in a single run) at all points $x_i$ of the design area, the gradient of the objective function $g^{(t)}(x_i)$.  This computation, conceptually, consists in considering that fictitious currents are induced in the the structure when transits from a state denoted $old$ to a state labelled $new$. This transition may be due to an evolution of the geometric or physical parameters of the system. The gradient is then used to modify the design variable values at each voxel of the design domain, in order to increase the FOM. The algorithm is performed iteratively, by pushing the  continuous profile towards a  discrete profile. This is accomplished through filtering (blurring) and projection (binarization) schemes in each step. A more detailed technical discussion on the computation of the gradient of the figure of merit applied here is provided in reference \cite{Kofi_TO}. 
%The algorithm starts  by  generating a couple of profiles termed father (or trend) and mother (or fluctuations). 
For those who are not familiar with the FTA concept, it may be useful to clarify the underlying motivations and intuitions guiding  the choice of terms: mother profile and father profile.
Father and Mother are terms commonly used in the theory of signal processing by the wavelet transform. Wavelets are defined by a wavelet function also called  "mother" wavelet and a scaling function also called "father" wavelet. The wavelet function is  a   high band-pass filter (fluctuating function) while the scaling function filters the lowest level (scaling function). In the FTA, the couple of initial geometries have the same aspect and plays the same role as the father and mother wavelets used in signal processing. Besides, they handle the mean features of the final optimized structure:  father geometry handles the global aspect (Trend or scale) while the mother profile holds the minimun size  features. Now that this point is clarified, let's recall briefly the basic idea  of the FTA scheme.  
During the first iteration, the mother profile becomes the old profile while the father profile is the new "trend". 
Only a forward simulation is performed on the father/new profile while a forward and  adjoint computations  (both computations  are obtained with only one simulation) are performed on the mother/old profile. Hence a new profile is updated thanks to  gradient based-like algorithm. The variable to be optimized, can undergo an ascending or descending increment as explained in reference \cite{Kofi_TO}.  At each iteration, and for each point of the design region, only the  increment direction leading to the best result is kept.
The ability of a gradient-based TO strategy, to yield  satisfactory results, strongly depends on the initial conditions of the gradient-based algorithm. These initial geometries must be carefully selected to avoid undesirable local minima. 
In \cite{Kofi_TO}, while designing a 1D functional metasurface, a cosine function
\begin{equation}\label{fathercosine}
    \rho_{trend}(x)=\dfrac{1}{2}\left[cos\left(\dfrac{2\pi \eta }{d}x\right)+1\right],
\end{equation}
has been identified to be an optimal father profile allowing to efficiently target  a basin of local  minima.
This  profile function depends on a parameter $\eta$ which controls, both the maximum/ minimum feature sizes, and also the maximum number of  nanorods of which the final binarized 1D metasurface is made up. The final optimized result  may be, and in certain case, is, very sensitive to the variations of this parameter $\eta$. While a judicious choice of the father profile parameter $\eta$, systematically leads to a very rapid convergence towards a local minimum of the figure of merit, and this, independently of the mother function, a non judicious choice of $\eta$ may drive the  algorithm  towards  basins of undesired solutions.
The instability introduced by the very high sensitivity of the FTA algorithm to the parameter $\eta$ strongly penalizes the method.
A trivial solution allowing to face this drawback, consists in randomly sweeping $\eta$-values on a given interval. However this randomized process can have a prohibitive cost in terms of computation time. The question is how to generate better initial candidates,  preserving the advantages introduced by the use of the initial sinusoidal father profile function.
%of finding an efficient, robust and inexpensive computation time method, which allows, while overcoming the instability linked to the parameter $\eta$, to preserve the advantages introduced by the use of %Il est donc important de cibler toute une famille de profil
The underlying idea of the solution we suggest is based on the use of the DSA in order to generate, from a set of initial  best geometries, a basin of fittest individuals. In the next section, the basic sketch  of the DSA  is introduced.
%%%%%%%%%%%%%%%%%%%%%%%%%%  body  %%%%%%%%%%%%%%%%%%%%%%%%%%
\subsection{ The fluctuations-trend analysis (FTA)  assisted by the diamond-square algorithm (DSA): FTA-DSA}
%%%%%%%%%%%%%%%%%%%%%%%%%%  body 
%\subsection{Diamond-Square Algorithm DSA}
For the optimization problem under consideration: designing a 1D metagrating capable of deflecting an incident wave to a given direction, we use the FTA to find an optimal permittivity function $\varepsilon (x)$. In practice, the FTA is performed iteratively using a gradient-like method making it very sensitive to the initial conditions; which poses a problem of robustness.  A trivial solution to this drawback, consists in randomly  checking many initial profiles but, as is highlighted previously, this random process can have a prohibitive cost in terms of computation time. 

%%%%%%%%%%%%%%%%%%%%%%%%%%
\begin{figure}[htb!]
 \centering
         {\includegraphics[width=.5\textwidth]{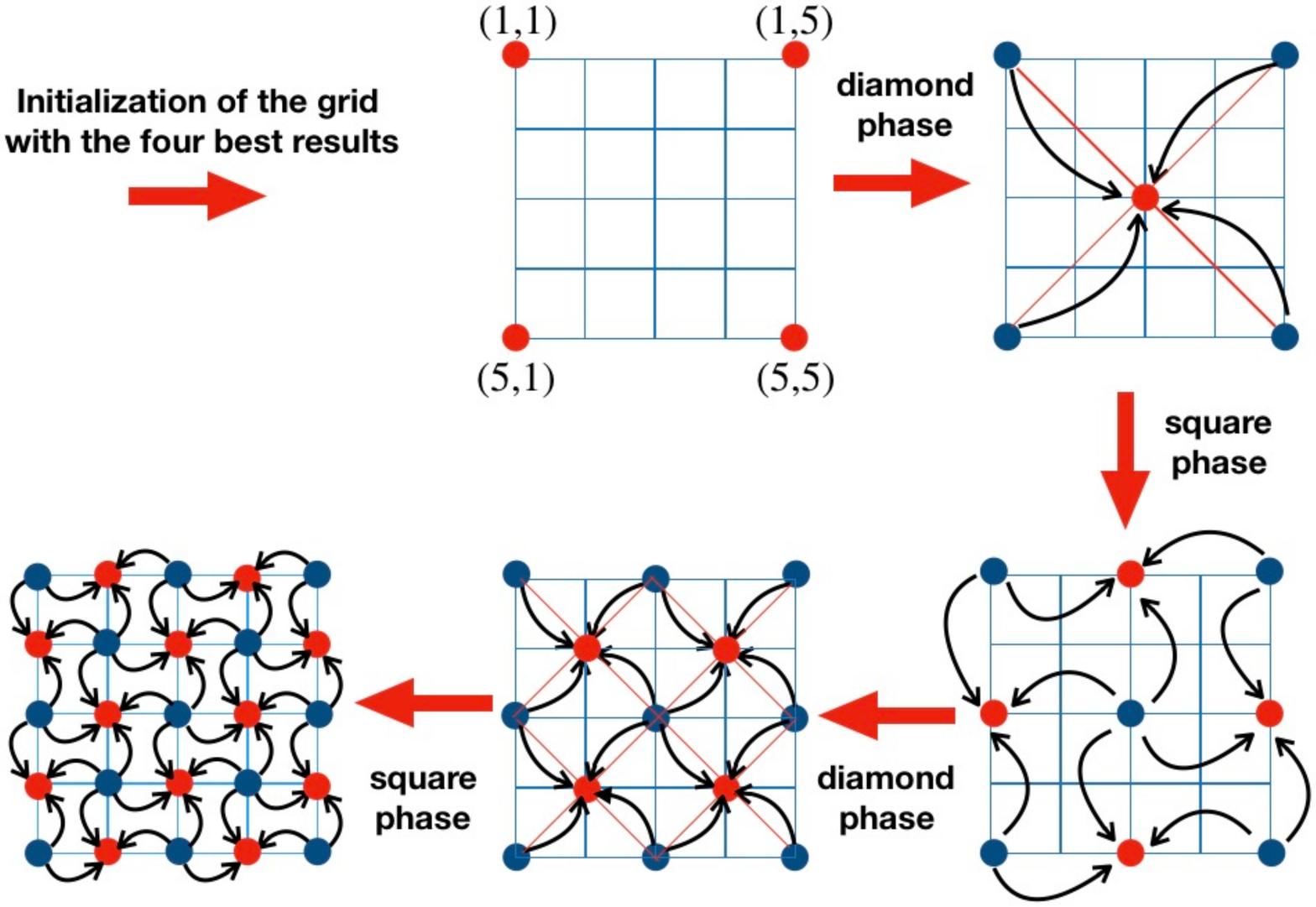}}
        %\caption{\label{deflection_GG_high} toto}
\caption{\label{sketch_SQM3}
Flowchart of the DSA performed on  a $5\times 5$ 2D grid. Here $n$ is set to $n=3$, which means, that, after the initialization of the four corners of the grid, $2$ cycles of diamond-square phases  are implemented.}
\end{figure}

\begin{figure}[htb!]
 \centering
         {\includegraphics[width=.4\textwidth]{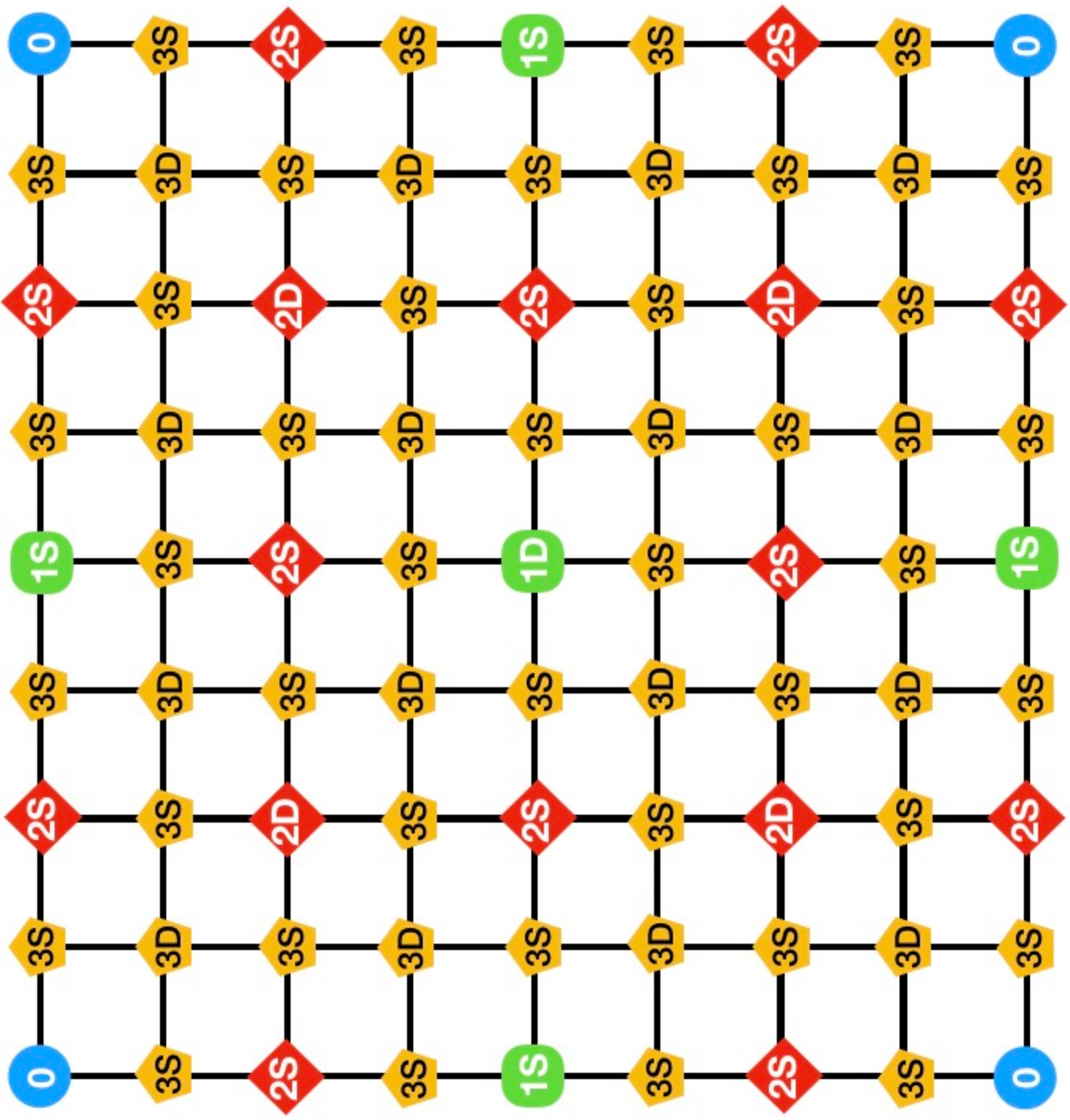}}
        %\caption{\label{deflection_GG_high} toto}
\caption{\label{sketch_SQM4}
Architecture of the 2D DSA grid obtained for $n=4$ yielding a $9 \times 9$ 2D grid. $3$ cycles are performed. A node obtained at the end of the  $Nth$ cycle is denoted $Nd$ if it comes from a diamond phase and $Ns$ if it is  generated by a square phase.}
\end{figure}
%%%%%%%%%%%%%%%%%%%%%%%%%%

Our idea is to perform the FTA with a small number of iterations using some initial profiles identified as the best candidates at a given iteration. At the beginning, we depart from a certain set of random candidates, run the FTA with a moderate number of iterations and select the four best among them. Then we use the DSA (the principle of which is given in the following) to generate an offspring of the latter and then run the FTA process over till a criterion (a given efficiency in a given diffraction direction) is satisfied. Thus the main role of the DSA ,here, is to provide the FTA with a set of new generations with a sufficient variety yet holding the main characteristics of the four initial parents. For the readers not familiar with the DSA, we now provide the basic principles of this algorithm as applied in our problem. The diamond-square algorithm, firstly proposed by  Fournier, Fussell and Carpenter  \cite{Fournier}, is a method originally devised to generate  two-dimensional landscapes for computer graphics. The algorithm starts with the generation of a 2D grid of size $(2^{n-1}+1) \times (2^{n-1}+1)$, $n \neq 1$. At the beginning, the four corners of the 2D grid are initialized. In the optimization problem under consideration, we are interested in designing a permittivity function $\varepsilon(x)$ of a metagrating, at some points $x$ of the design area. Therefore the corners will be initialized by the four best candidates stemming from the first few FTA iterations. \\

After the initialization, two phases, namely {\it diamond phase} and {\it square phase}, are progressively and  alternately  performed until all values of the all nodes of the 2D grid have been set. The couple (diamond, square)  determines a cycle of the  DSA.  For a given value of the integer  $n$,  $n-1$  cycles labelled $p$, ($p \in \mathbf{N} \cap [2,n]$) are performed  to set all nodes of the 2D grid. In this paper $n$ will also denote  the order of the DSA. In the $p^{th}$ cycle, one considers all $2^{p-1} \times 2^{p-1}$,  ($p \in \mathbf{N} \cap [2,n]$), sub-squares. To the center of each of them $(I,J)$ , we associate the average of $\varepsilon (x)$ over the corners :

\begin{equation} \varepsilon^{(I,J)}(x)=\dfrac{1}{4} \sum_{i,j}\varepsilon^{(i,j)}(x) +\left(\dfrac{u}{2^{u}}\right)^p (2r-1)\end{equation} 
where $u\in[0,1]$ is a constant and $r$ a random number in the range $[0,1]$. The sum is performed on the value of the functions at the four closest nodes. 

This phase is termed diamond because, drawing lines from the initial corners to the generated midpoint, leads to a diamond pattern. In the square phase, one considers all diamond-kind patterns made of the middle points computed in the previous diamond phase and the four previous square corners.  The middle point of this diamond kind pattern is then  computed as the average of these  four nodes. The name square phase comes from the fact that by drawing lines from the corners to the midpoint one creates a square pattern. An illustration is given in fig. \ref{sketch_SQM3} where $n$ is set to ${\color[rgb]{1,0,0}{3}}$, which leads to a $5\times 5$ 2D grid. After the initialization of the four corners of the grid, two rounds of diamond-square steps  are implemented. We present in fig. \ref{sketch_SQM4} the architecture of a 2D grid obtained from  a higher order of DSA. In this figure, $n=4$ yielding a $9 \times 9$ 2D grid with $81$ nodes. In order to facilitate the reading of the grid and for the sake of convenience, we adopt the following notation: a node obtained at the end of the  $Nth$ cycle is denoted $ND$ if it comes from a diamond phase and $NS$ if it is  generated by a square phase. These nodes can also be identified by their coordinates $(i,j)$ in the rectangular mesh.

\begin{figure}[htb!]
 \centering
        {\includegraphics[width=.5\textwidth]{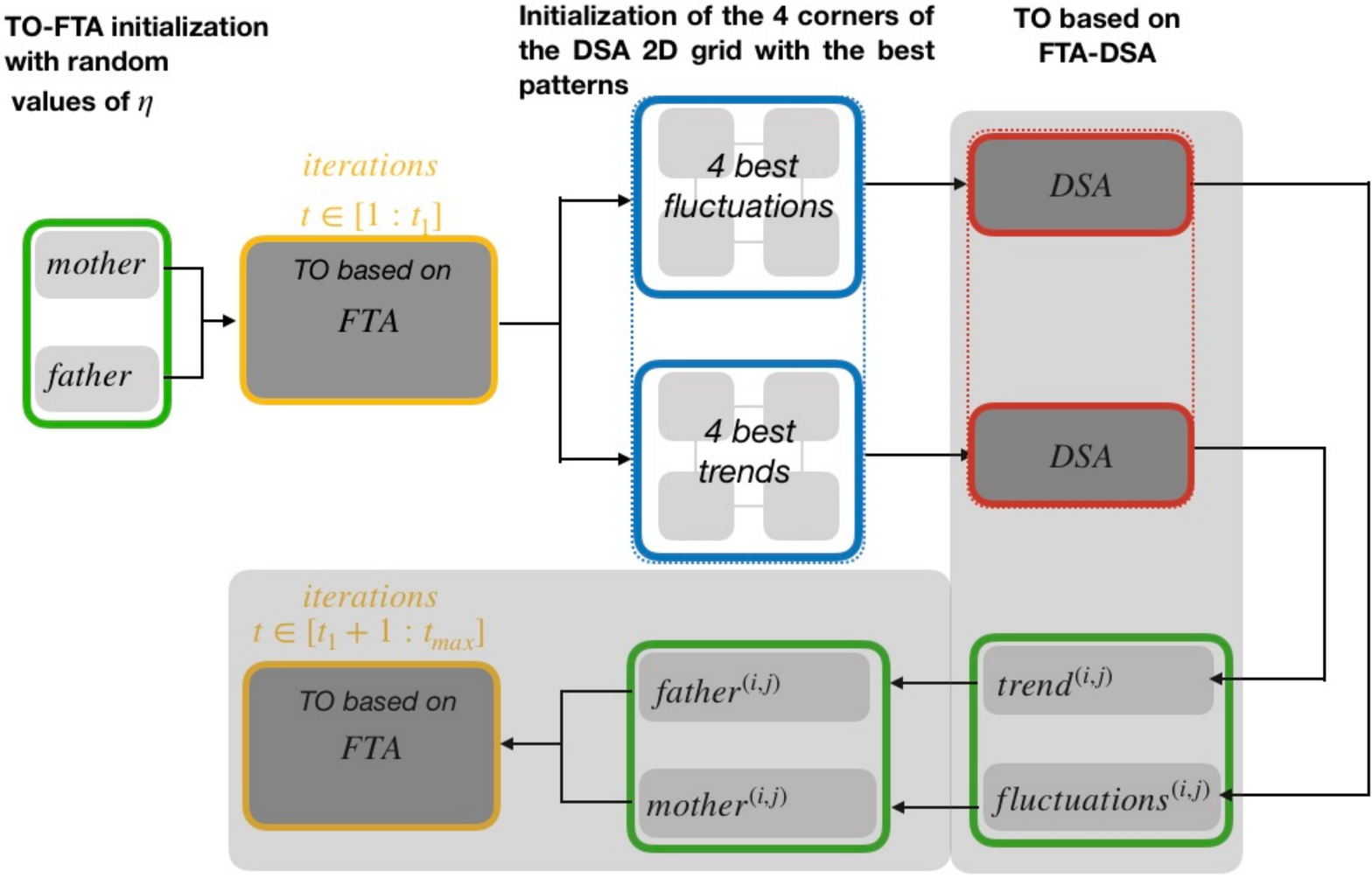}}
\caption{\label{sketch_SQM}
Flowchart of the topology optimization based on FTA-DSA.}
\end{figure}

Equipped with these coupled algorithms, we are now ready to handle our design problem which is done in the next section.

%%%%%%%%%%%%%%%%%%%%%%%%%%%%%%%%%
\section{Results}
%%%%%%%%%%%%%%%%%%%%%%%%%%%%%%%%%
%%%%%%%%%%%%%%%%%%%%%%%%%%%%%%%%%
\begin{figure}[htb!]
 \centering
 %\subfigure [\label{geometry_1D} 1D metagrating]
 {\includegraphics[width=0.45\textwidth]{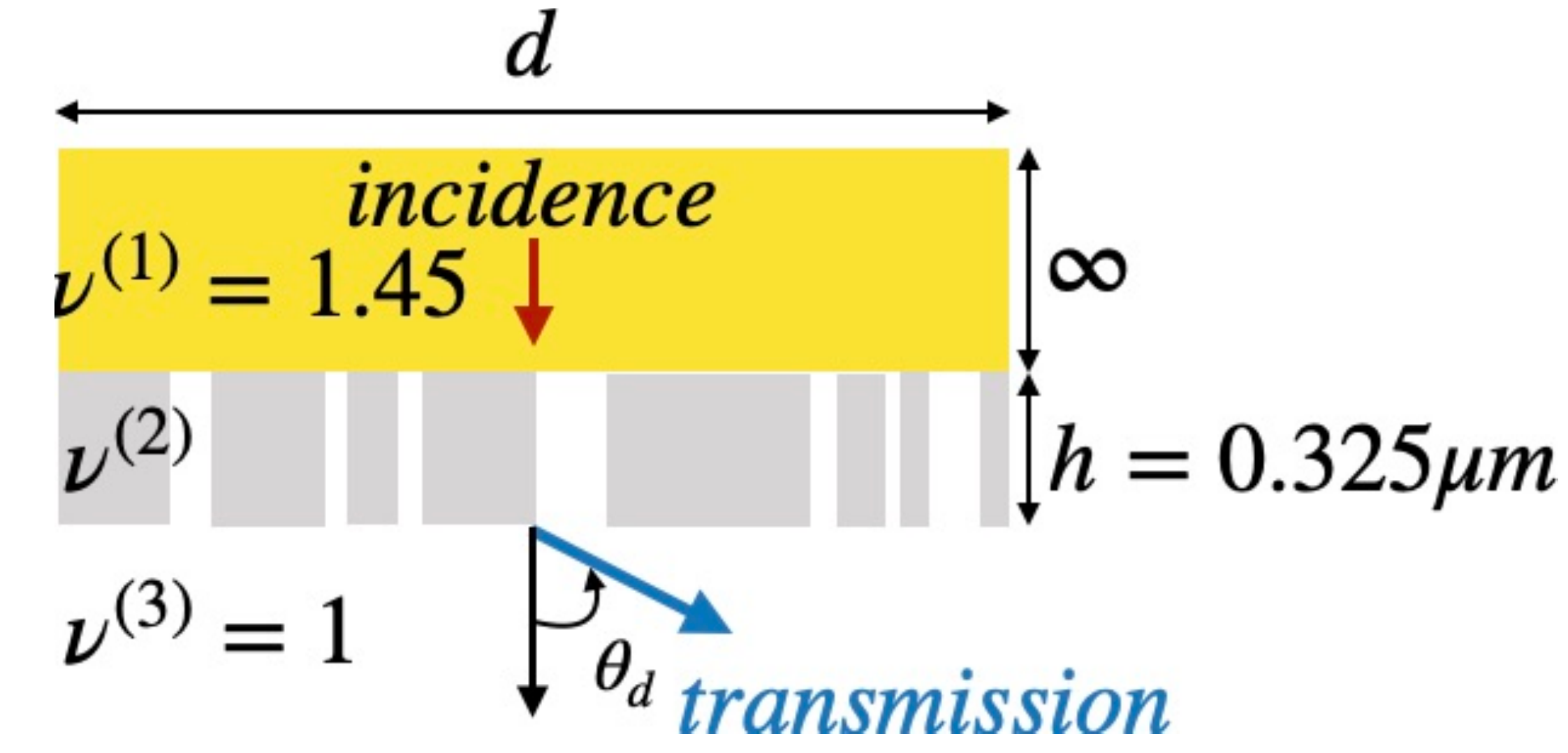}}
\caption{\label{geometry_1D} 1D metagrating  Geometry of a 1D metagrating deflecting a TM-polarized normal incident plane wave into $\theta_d$ angle}
\end{figure}

\begin{figure}[htb!]
 \centering
   \subfigure [\label{sketch_FTA_DSA_deflector} deflector configuration]
        {\includegraphics[width=0.2\textwidth]{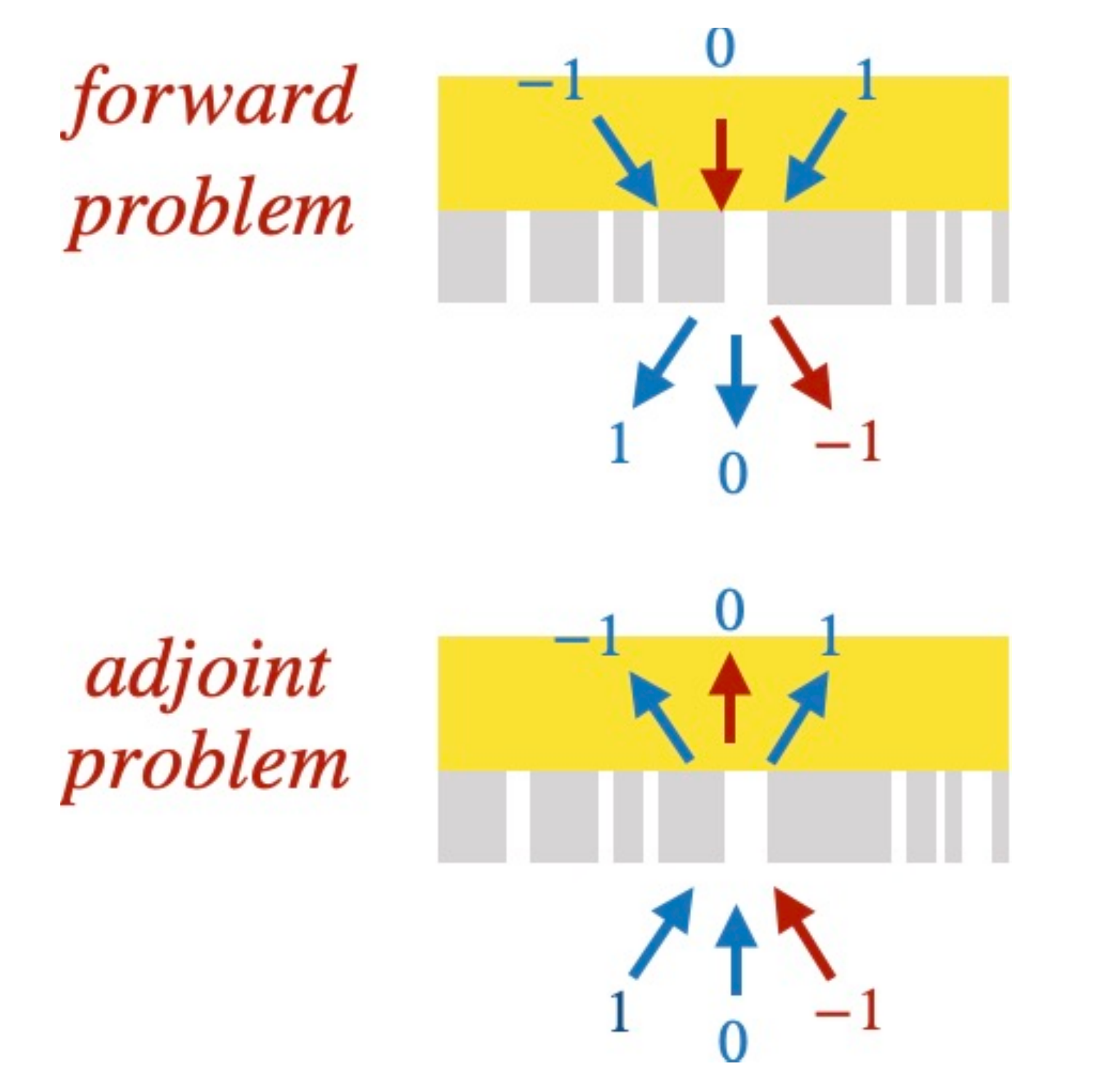}}
        \hspace{0.5cm} \vline \hspace{0.5cm}       
 \centering
   \subfigure [\label{sketch_FTA_DSA_neg_refraction} Anomalous negative refraction]
        {\includegraphics[width=0.15\textwidth]{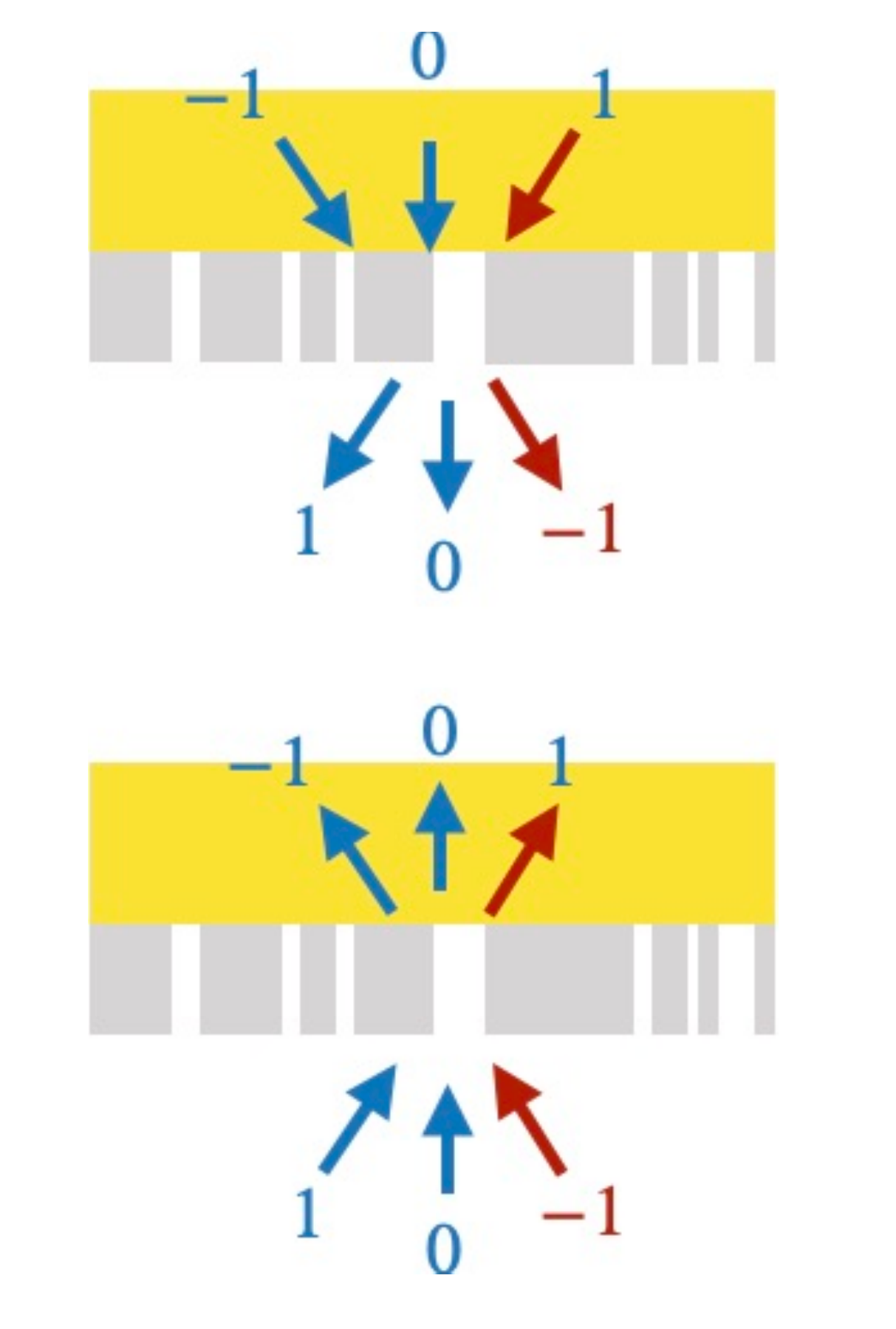}}
     
\caption{\label{deflection_GG}
Sketch of the direct and adjoint simulations used in the topology optimization of metagratings. Figures \ref{sketch_FTA_DSA_deflector} and \ref{sketch_FTA_DSA_neg_refraction} present the direct  and the adjoint computations methods for the deflector and anomalous refraction configurations respectively.}
\end{figure}
%%%%%%%%%%%%%%%%%%%%%%%%%%%%%%%%%
We apply the proposed algorithm to the inverse design of 1D metagratings earlier studied in \cite{Yang1, Jiang}, that deflect a normally-incident TM-polarized plane wave, with wavelenght $\lambda$ onto a particular transmitted angle $\theta_d$. The optimized device consists of Si-nanorods with refraction index $\nu^{(2)} = 3.6082$, deposited on a $SiO_2$ substrate ($\nu^{(1)} = 1.45$). The  grating's height is set to $h=0.325 \mu m$. See Fig. \ref{geometry_1D}. 
In order to get the best devices allowing to initialize the DSA grid, we first generate a set of random values of the parameter $\eta$ as follows: 
\begin{equation}\label{eta_generation}
    \eta\in\left\{(\eta_{max}-\eta_{min})rand(1,N_{\eta})+\eta_{min} \right\},
\end{equation}
where $N_{\eta}$ denotes the number of father profiles used as initial geometries in the first round of the  FTA. The parameter $N_{\eta}$ is set to $N_{\eta}=10$, $\eta_{max}=5$ and $\eta_{min}=3.5$. For each father profile, the  fluctuation or mother function is  generated through a  Druden-Vesecky ocean bandlimited spectrum \cite{Valenzuela, Wetzel, Apel, DV, Yueh1997, Yueh_1994}. 
Table \ref{tab_old_last_mother} presents results obtained on a $5 \times 5$ diamond-square 2D array.  Two  values of the incident field wavelength namely,  $\lambda=0.9\mu m$ and $\lambda=1.1\mu m$ are investigated. The four initial parents are located at the four nodes $(1,1)$, $(1,5)$, $(5,1)$ and $(5,5)$ while all the other nodes correspond to the offspring results. As shown in this table satisfactorily high efficiencies are reached.  Here five iterations are used in the first part of the FTA and $50$ iterations are performed in the second round of the FTA. As expected, for the chosen incident wavelength,  a large part of offspring leads to  better fitness than parents. For $\lambda=0.9\mu m$,  offspring of node $(2,5)$  yields  slightly better than the best father $\varepsilon^{(1,1)}$,
while $\varepsilon^{(2,4)}$ is the best result for $\lambda=1.1 \mu m$.   $\varepsilon^{(i,j)}$ denotes the optimized permittivity profile associated with the nodes $(i,j)$ of the DSA landscape.   
\begin{table}[htb]
\centering
\begin{tabular}{|c||c c c c c|}
%\multicolumn{6}{c}{} \\
\hline %\hline
%{\tiny{
{j $\rightarrow$} & 1 & 2 & 3  &  4 &  5\\
\hline
\hline
{i $\downarrow$}  &  &  & $\lambda=0.9\mu m$ &  &  \\
\hline
1 &  {\color{red}0.9736} &   0.8925 &  0.9605 &   0.9695 &   {\color{red}0.9688}\\
2 &  0.9624 &   0.9660 &  0.9616 &   0.9657 &   {\color{blue}0.9742}\\
3 &  0.8898 &   0.8930 &  0.8880 &   0.5886 &   0.6708\\
4 &  0.8963 &   0.8927 &  0.1979 &   0.6867 &   0.6210\\
5 &  {\color{red}0.8885} &   0.8890 &  0.3820 &   0.5845 &   {\color{red}0.9540}\\
\hline
\hline
{i $\downarrow$} &  &  & $\lambda=1.1\mu m$ &  &  \\
\hline
1    &  {\color{red}0.7806} &   0.7714  &  0.7775 &  0.7775                &  {\color{red}0.7492}\\
2    &           0.7714     &   0.7714  &  0.7806 &  {\color{blue}0.7825}  &  0.7775\\
3    &  0.7714 &   0.7806  &  0.7678 &  0.7678  &  0.7621\\
4    &  0.7678 &   0.7714  &  0.7714 &  0.7678  &  0.7678\\
5    &  {\color{red}0.7652} &   0.7714  &  0.7678 &  0.7678  &  {\color{red}0.7621}\\
\hline
\end{tabular}
\caption{\label{tab_old_last_mother}  Panel of efficiencies of a 1D dielectric metasurface obtained by the  FTA-DSA on a $5 \times 5$ 2D grid for two values of wavelength: $\lambda=0.9\mu m$  and $\lambda=1.1\mu m$. The structure is optimized to deflect a normally TM polarized incident plane wave onto $\theta_d= 60^o$. Numerical parameters: $d=\sqrt{\varepsilon_3}/sin(\theta_d)$, 
$t_{1}=5$, $t_{max}=55$, $\eta_{min}=3.5$, $\eta_{max}=5$, $ N_{\eta}=10$.}
\end{table} 

\begin{figure}[htb]%[htb!]
%=======================================
%\hline
%\centering
%\vspace{0.5cm}
% $\lambda=0.9\mu m$\\
\centering
   \subfigure [\label{epsilon_SQDM_11_theta_60_lambd09} $\varepsilon^{(1,1)}$, $\lambda=0.9\mu m$]
        {\includegraphics[width=0.22\textwidth]{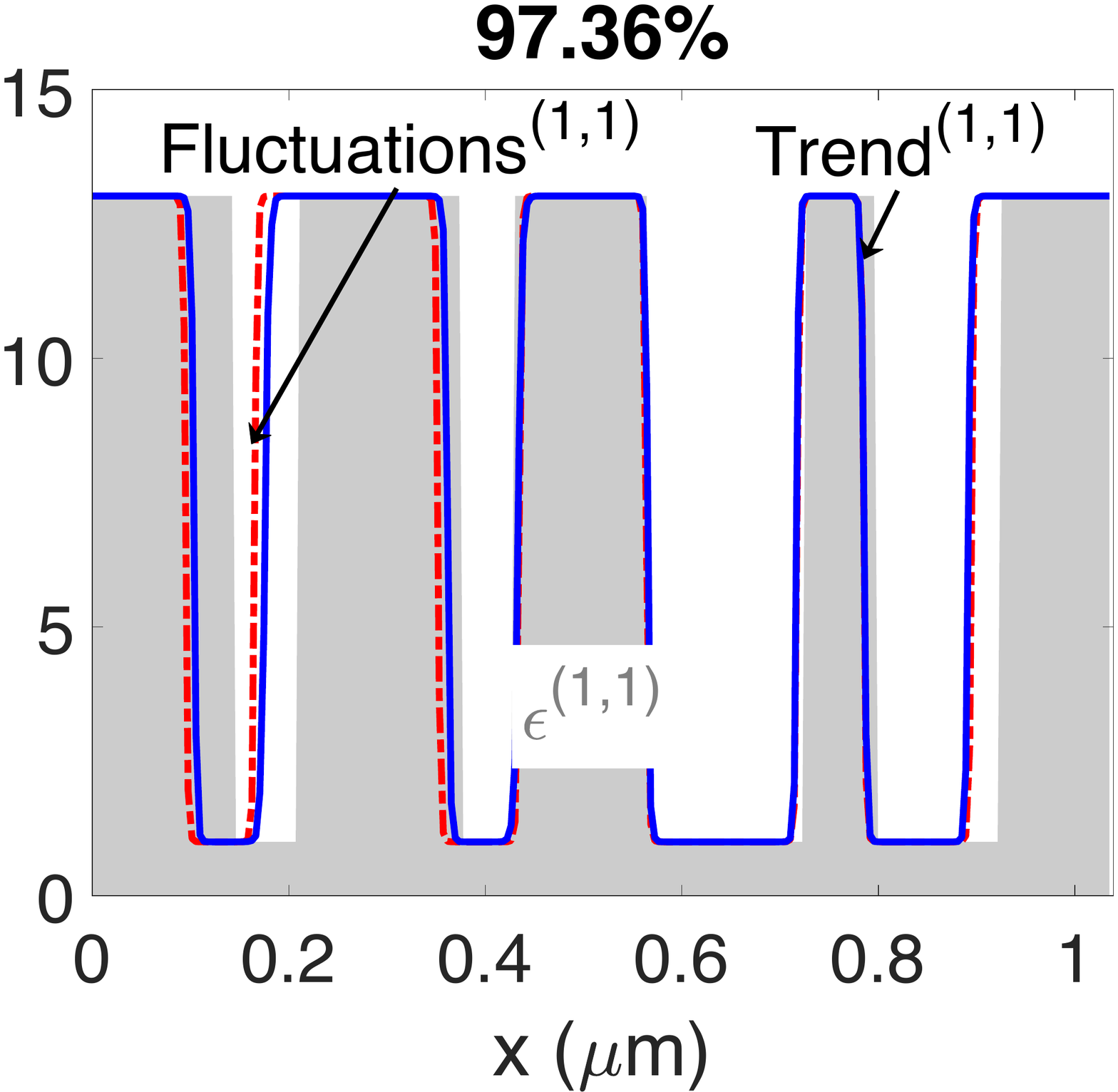}}
 %\centering
 %  \subfigure [\label{epsilon_SQDM_22_theta_60_lambd09} $\varepsilon^{(2,2)}$]
 %       {\includegraphics[width=0.32\textwidth]{epsilon_SQDM_22_theta_60_best_int.eps}}
\centering
   \subfigure [\label{epsilon_SQDM_25_theta_60_lambd09} $\varepsilon^{(2,5)}$, $\lambda=0.9\mu m$]
        {\includegraphics[width=0.22\textwidth]{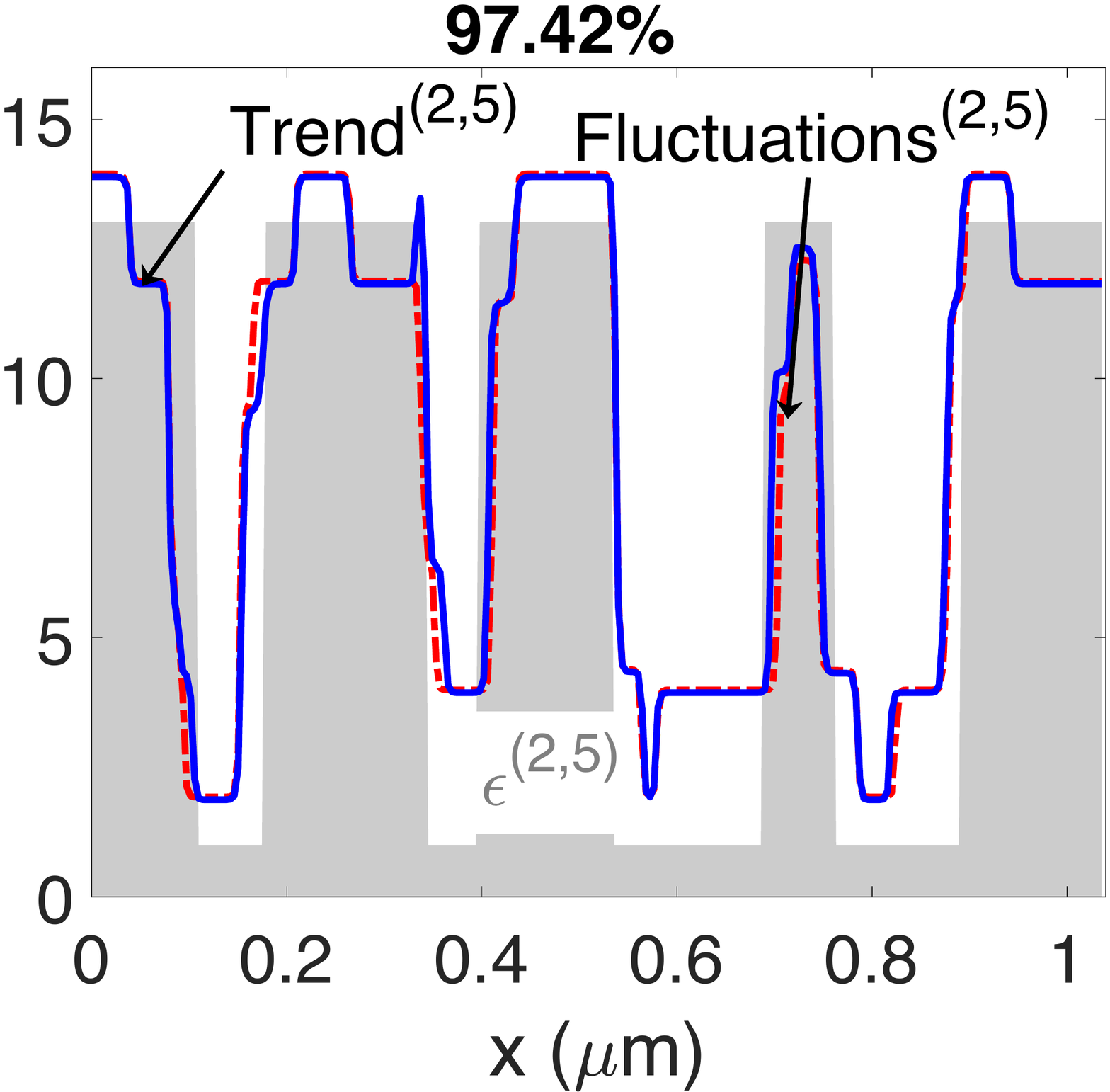}}
%=======================================
%\hline
%\vspace{0.5cm}
%\centering
 %$\lambda=1.1\mu m$\\
 \centering
 \subfigure [\label{epsilon_SQDM_lambd11_11_theta_60_lambd11} $\varepsilon^{(1,1)}$, $\lambda=1.1\mu m$]
        {\includegraphics[width=0.22\textwidth]{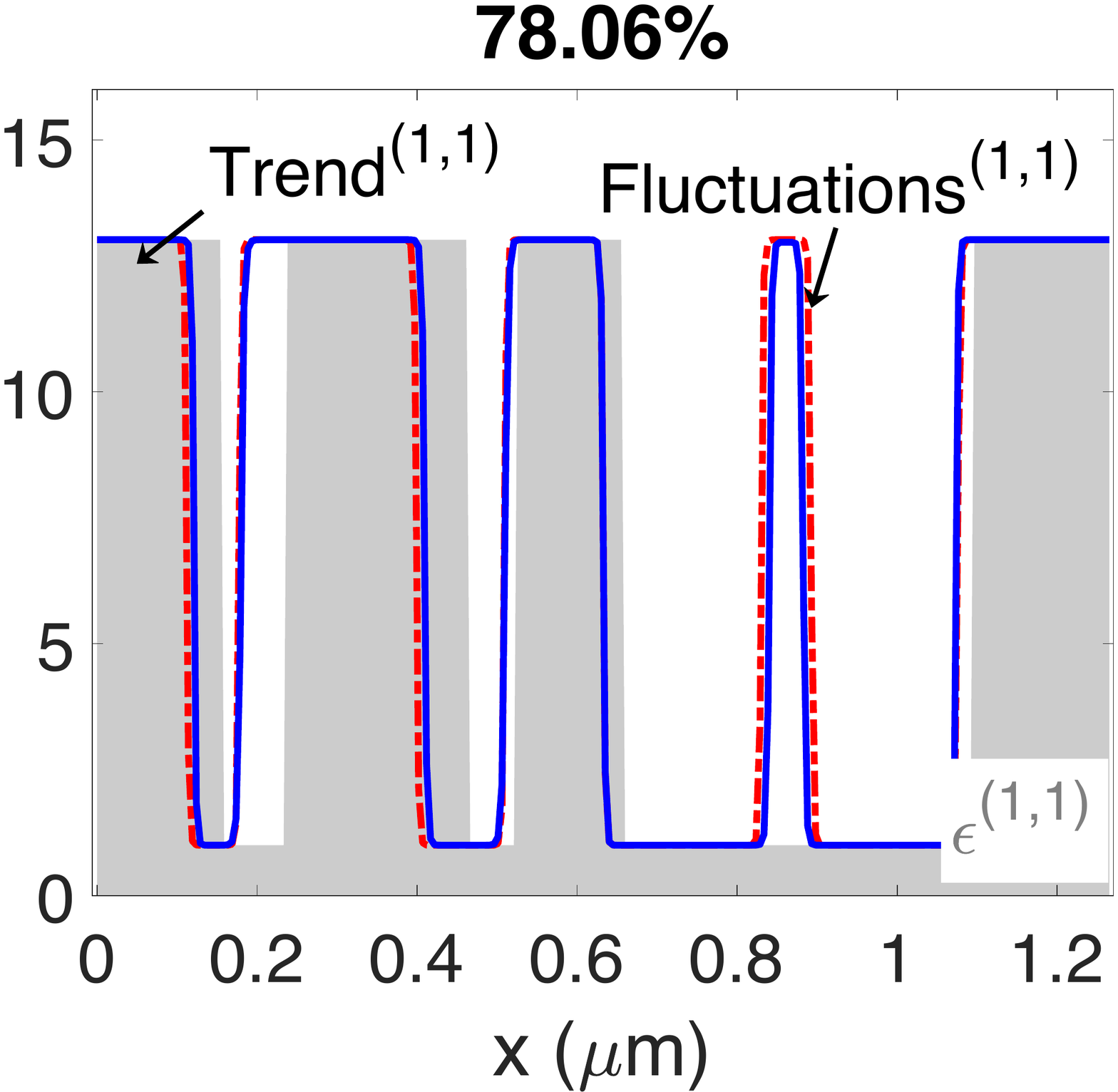}}
        %\caption{\label{deflection_GG_high} toto}
% \centering
% \subfigure [\label{epsilon_SQDM_lambd11_13_theta_60_lambd11} $\varepsilon^{(1,3)}$]
% {\includegraphics[width=0.32\textwidth]{epsilon_SQDM_lambd11_13_theta_60_best_int.eps}}
 \centering
 \subfigure [\label{epsilon_SQDM_lambd11_24_theta_60_lambd11} $\varepsilon^{(2,4)}$, $\lambda=1.1\mu m$]
        {\includegraphics[width=0.22\textwidth]{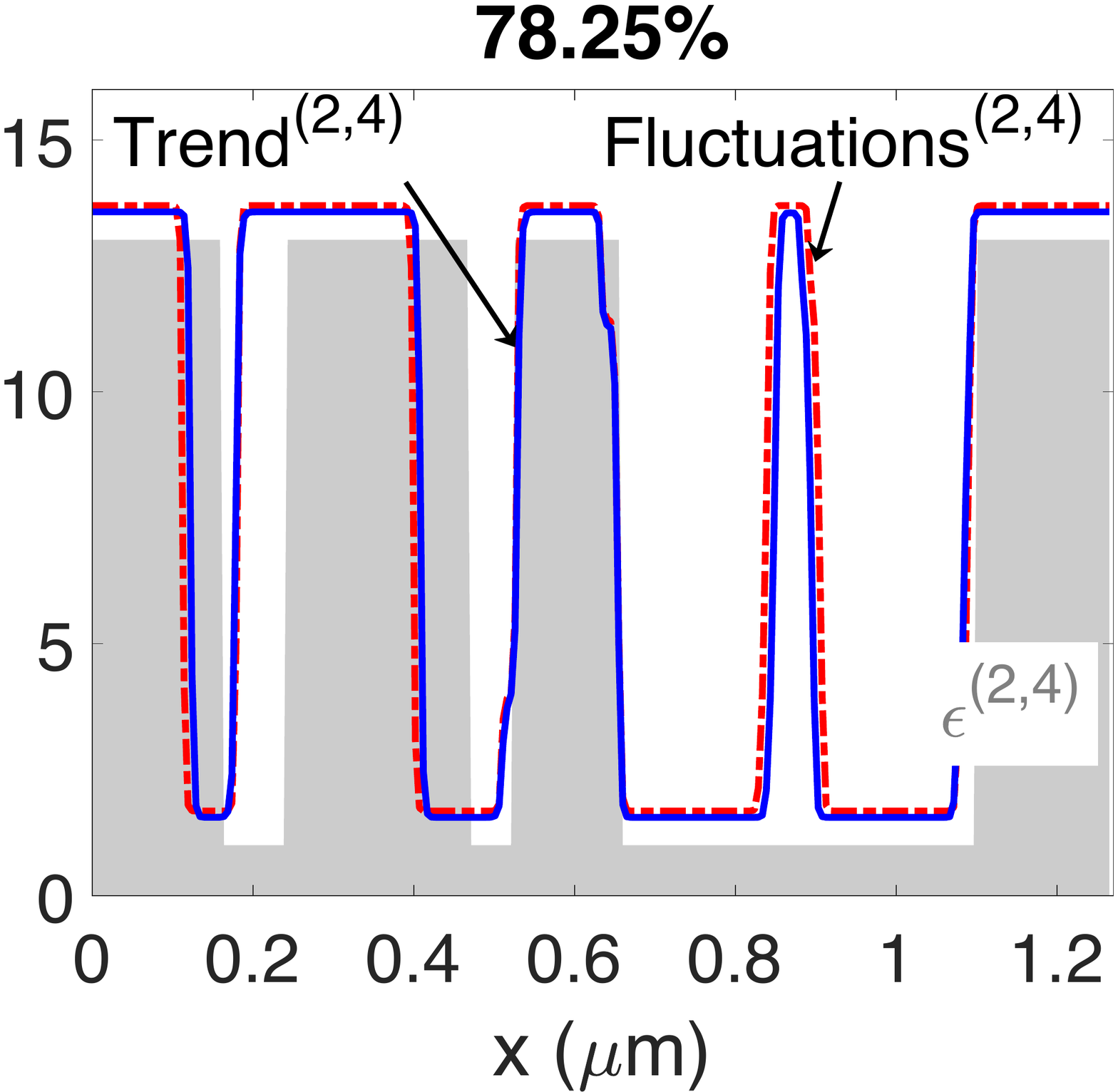}}
%\hline
%=======================================
%\centering
%   \subfigure [\label{high_trans} $\varepsilon^{(4,3)}$]
%        {\includegraphics[width=0.45\textwidth]{epsilon_SQDM_lambd11_43_theta_60_best_int.eps}}
\caption{\label{epsilon_SQDM_theta_60_best_tab1}
Sketch of two final optimized devices for two wavelengths $\lambda=0.9 \mu m$ (figs. \ref{epsilon_SQDM_11_theta_60_lambd09}, \ref{epsilon_SQDM_25_theta_60_lambd09}), 
and $\lambda=1.1\mu m$ (figs. \ref{epsilon_SQDM_lambd11_11_theta_60_lambd11},  \ref{epsilon_SQDM_lambd11_24_theta_60_lambd11}). The obtained optimized profile and the two initial geometries namely Trend and fluctuations profiles are displayed in each case. Numerical parameters: $\lambda=0.9\mu m$, deflection angle $\theta_{d}=60^{\circ}$, $t_1=5$, $N_{eta}=5$, $\eta_{min}=3.5$, $\eta_{max}=5$,  grating period  $d=\nu_3\lambda/sin(\theta_{d})$.}
\end{figure}
\begin{figure}[htb]%[htb!]
 \centering
      {\includegraphics[width=0.35\textwidth]{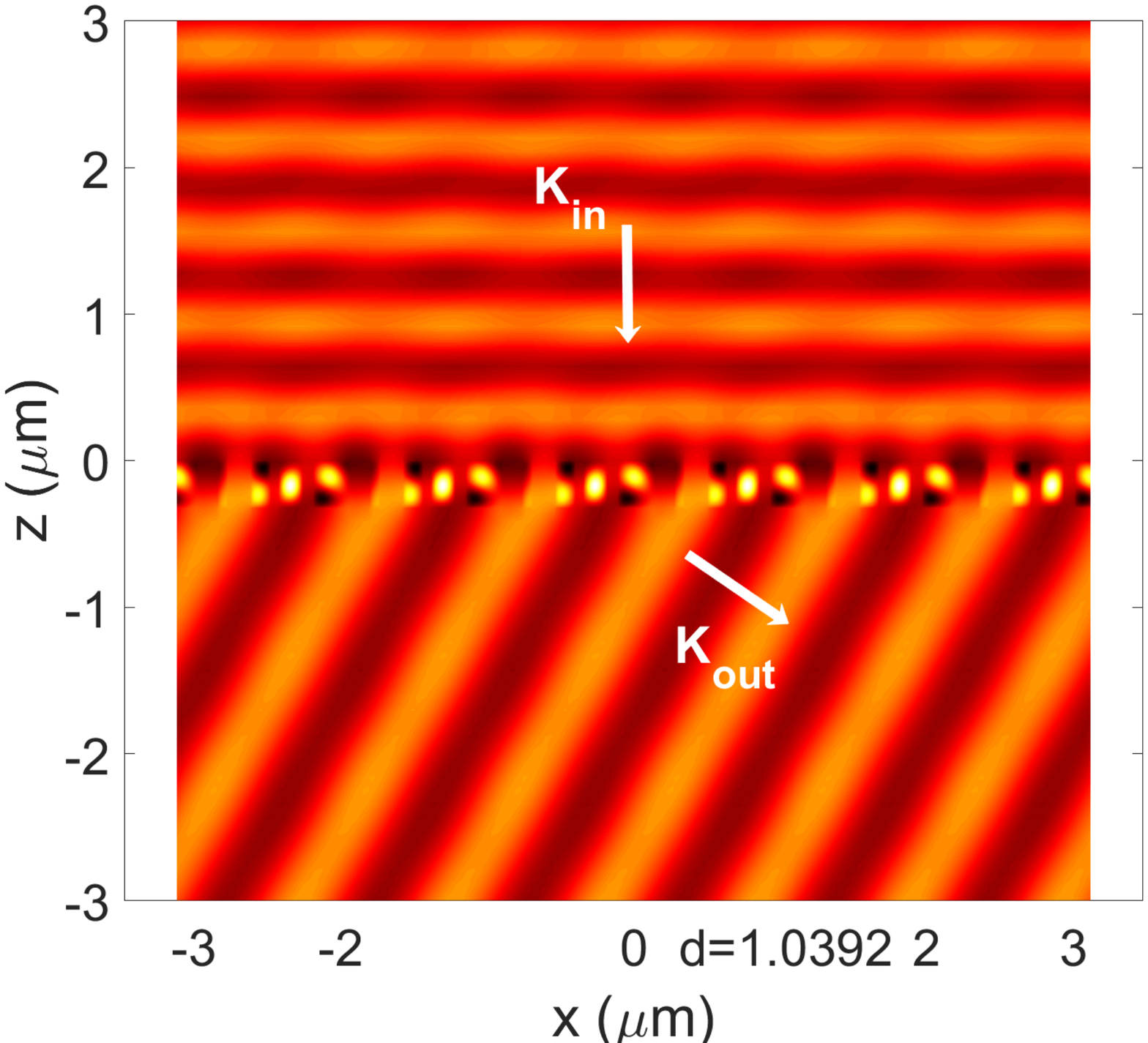}}
\caption{\label{filed_map_Hy__deflec_60_lambda09}
TFA-DSA applied to design a 1D high-transmission deflection  metagrating. Real part of the magnetic field. Illustration of the quality of the  deflection phenomenon supported by one of the final highest-transmission devices ($\varepsilon^{(2,5)}$).  Numerical parameters: $\lambda=0.9\mu m$, deflection angle $\theta_{d}=60^{\circ}$, $t_1=5$, $N_{eta}=10$, $\eta_{min}=5$, $\eta_{max}=3.5$,  grating period  $d=\nu_3\lambda/sin(\theta_{d})$. }
\end{figure}
\begin{figure}[htb]%[htb!]
 \centering
 \subfigure [\label{histogram_deflection_efficiency_lamb09_teta_60_deflec_FTA-DSA} $\lambda=0.9 \mu m$]
 {\includegraphics[width=0.22\textwidth]{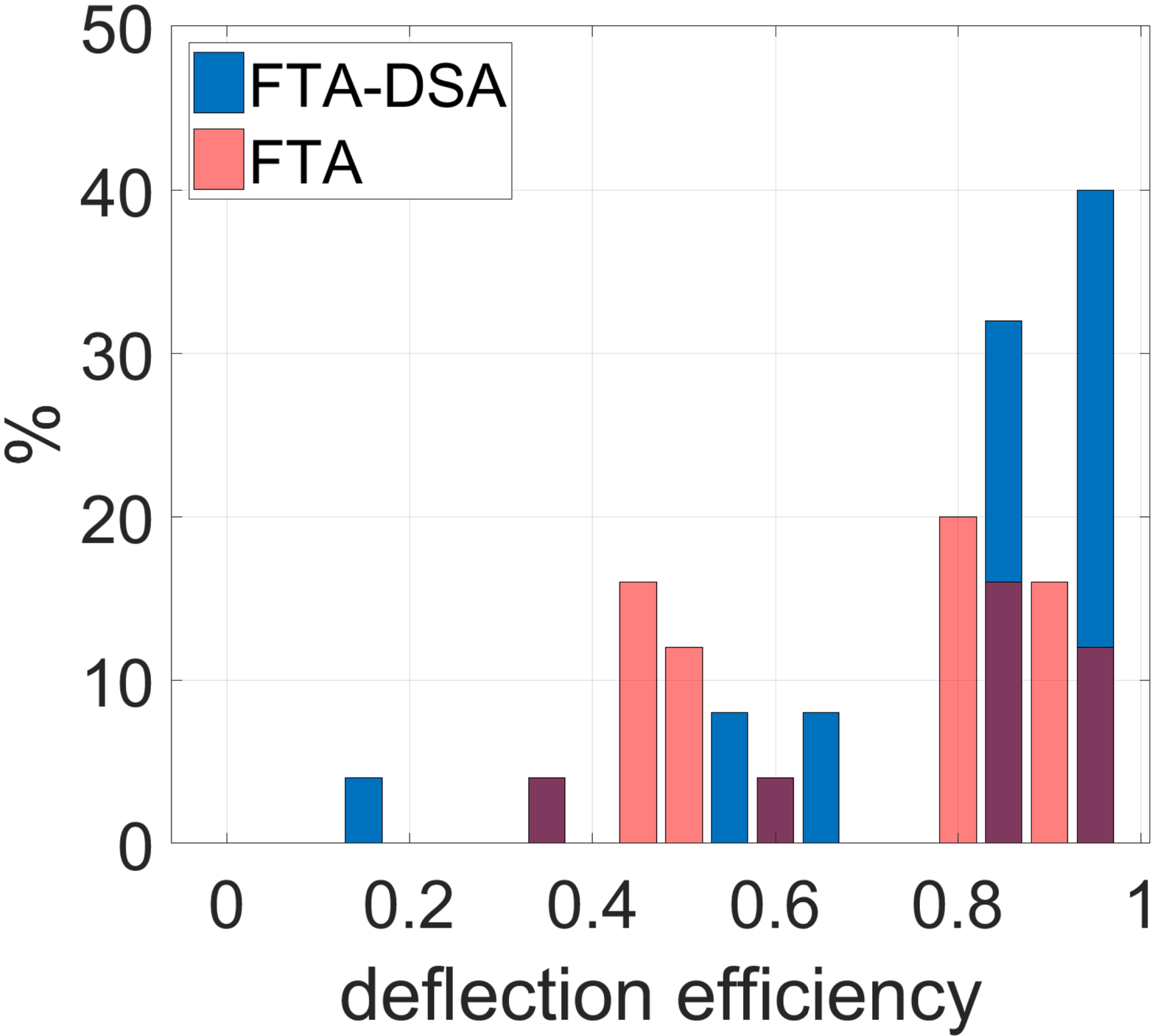}}
 \centering
   \subfigure [\label{histogram_deflection_efficiency_lamb11_teta_60_deflec_FTA-DSA} $\lambda=1.1 \mu m$]
        {\includegraphics[width=0.22\textwidth]{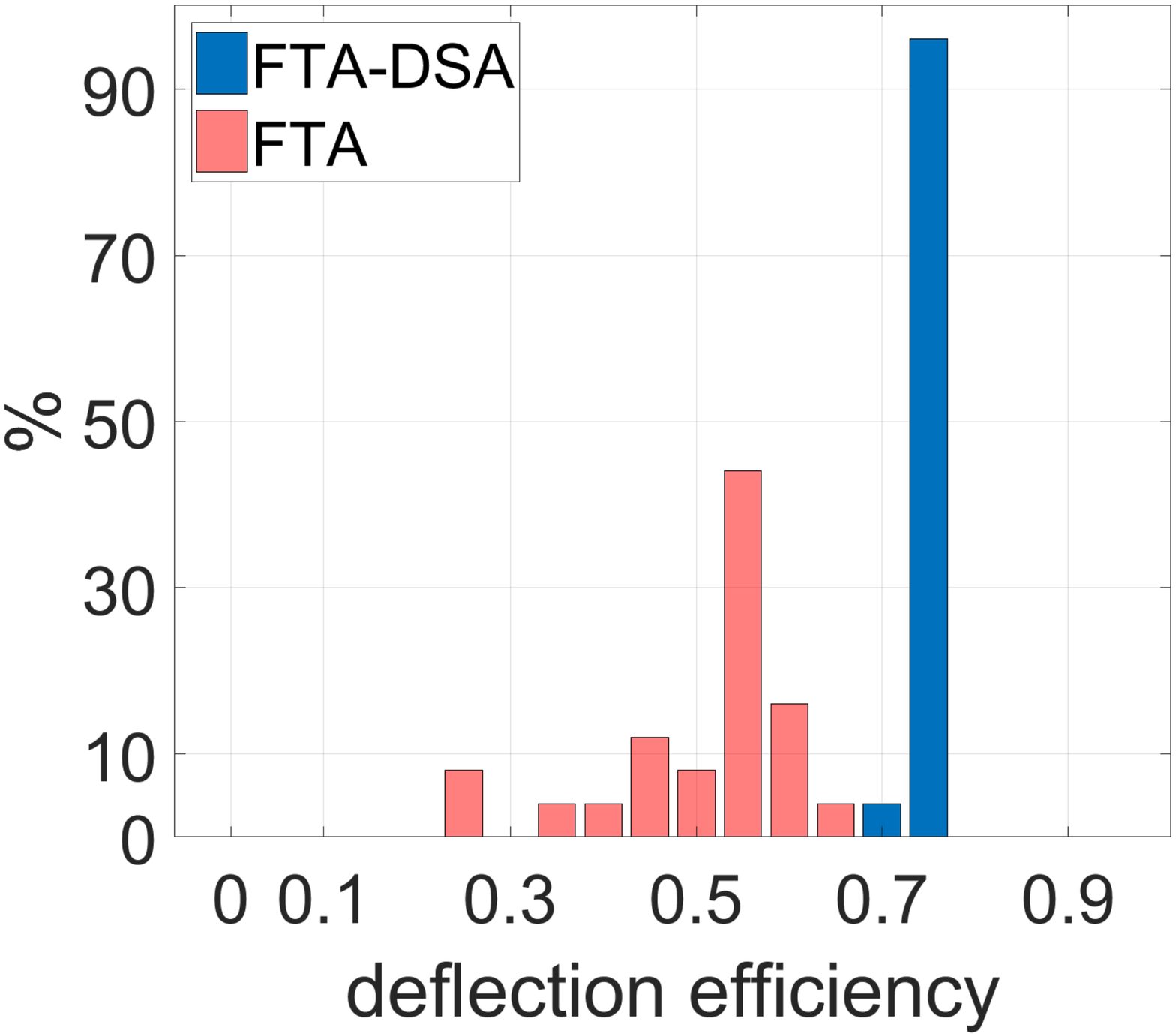}}
\caption{\label{histogram_deflection_efficiency_60}
Comparison of efficiencies histograms obtained with the FTA (ref color) and the FTA-DSA (blue) for $\lambda=0.9\mu m$ (Fig. \ref{histogram_deflection_efficiency_lamb09_teta_60_deflec_FTA-DSA}) and $\lambda=1.1\mu m$ (Fig. \ref{histogram_deflection_efficiency_lamb11_teta_60_deflec_FTA-DSA}).  Numerical parameters: deflection angle $\theta_{d}=60^{\circ}$, $t_1=5$, $N_{eta}=10$, $\eta_{min}=5$, $\eta_{max}=3.5$,  grating period  $d=\nu_3\lambda/sin(\theta_{d})$. }
\end{figure}

%%%%%%%%%%%%%%%%%%%%%%%%%%%%%%%%%%%%%%%%
Figures \ref{epsilon_SQDM_theta_60_best_tab1} present, some best optimized devices of the table \ref{tab_old_last_mother}. The initial father and mother profiles used in the second round of the TFA are also displayed in each case.
Figure \ref{epsilon_SQDM_11_theta_60_lambd09}, shows the result obtained for $\lambda=0.9 \mu m$ from the best parent. While fig. \ref{epsilon_SQDM_25_theta_60_lambd09} is related to initial elements obtained from inter-generational crosses.
Recall that intergenerational cross breeding results in an  average between permittivity functions of initial individuals. This explains the heckled-like specific shape of their trend and fluctuations profiles used in the second round of the TFA. As predicted, all these second-round couples of initial geometries hold similar features and yield final high performance auto-similar devices. The above observations are hold for a longer wavelength $\lambda=1.1\mu m$. See figs. \ref{epsilon_SQDM_lambd11_11_theta_60_lambd11}, and \ref{epsilon_SQDM_lambd11_24_theta_60_lambd11}. From these results, it appears that  that  efficient devices operating with a couple $(\lambda=0.9\mu m, \theta_d=60^{o})$ are a 4-nanorods type while a 3-nanorods-kind is exhibited in the case of $(\lambda=1.1\mu m$ at the same deflection angle $\theta_d=60^{o})$. 
We compute, and plot in fig. \ref{filed_map_Hy__deflec_60_lambda09} the real part of the magnetic field  through the best optimized final device \textsl{i.e.} $\varepsilon^{(2,5)}$. The deflection of the normally incident plane wave is clearly highlighted.
In Figs. \ref{histogram_deflection_efficiency_60}, we compare the efficiency histograms of optimized devices computed with the FTA and FTA-DSA, for  $\lambda=0.9\mu m$ (fig. \ref{histogram_deflection_efficiency_lamb09_teta_60_deflec_FTA-DSA}) and $\lambda=1.1\mu m$ (fig. \ref{histogram_deflection_efficiency_lamb11_teta_60_deflec_FTA-DSA}) for $\lambda=1.1 \mu m$.  As shown in these figures, $68\%$ of the $25$ realizations have efficiencies higher than $88\%$, indicating that the  $\eta$-landscape is efficiently and broadly scanned for the chosen range namely $\eta \in[3.5,5]$. These histograms also demonstrate clearly that adding DSA to the classical FTA provides systematically better results.\\ 

\begin{table}[htb]
\centering
\begin{tabular}{|c||c c c c c|}
%\multicolumn{6}{c}{} \\
\hline %\hline
 {j $\rightarrow$} & 1 & 2 & 3  &  4 &  5\\
\hline
\hline
{i $\downarrow$}                        &                     &                      & $\lambda=0.9 \mu m$ &  & \\
\hline
1 & {\color{red}0.9427} & {\color{blue}0.9705} & {\color{blue}0.9705} & 0.9486 & {\color{red}0.9669}\\
2 & {\color{blue}0.9705} & {\color{blue}0.9705} & 0.3105 & 0.9658 & 0.3221\\
3 & 0.9666 & 0.9605 & 0.9605 & 0.3145 & 0.3021\\
4 & 0.9619 & 0.2999 & 0.3125 & 0.3021 & 0.2992\\
5 & {\color{red}0.9568} & 0.2788 & 0.3218 & 0.3009 & {\color{red}0.9669}\\
\hline
\hline
{i $\downarrow$}  &                      &       &   $\lambda=1.1 \mu m$ &  & \\
\hline
1 & {\color{red}0.7735} & 0.7923 & 0.7918 & 0.7937 & {\color{red}0.7735}\\
2 & 0.7787 & 0.7890 & {\color{blue}{0.7982}} & 0.7777 & 0.7937\\
3 & 0.7890 & 0.7890 & 0.7726 & 0.7876 & 0.7872\\
4 & 0.7809 & 0.7809 & 0.7876 & 0.7746 & 0.7872\\
5 & {\color{red}0.7565} & 0.7934 & 0.7809 & 0.7872 & {\color{red}0.7727}\\
\hline
\end{tabular}
\caption{\label{tab_multi_level}   Panel of efficiencies of a 1D dielectric metasurface obtained by a TO based on a 10 layers-supervised/multi-layers  FTA-DSA-FTA.  The algorithm is performed  on a $5 \times 5$ 2D grid for two values of wavelength: $\lambda=0.9\mu m$ and $\lambda=1.1\mu m$. The structure is optimized to deflect a normally TM polarized incident plane wave onto $\theta_d= 60^o$. Numerical parameters: $d=\sqrt{\varepsilon_3}/sin(\theta_d)$, $t_{1}=5$, $t_{max}=55$, $\eta_{min}=3.5$, $\eta_{max}=5$, $ N_{\eta}=10$.}
\end{table} 
%=================================
\begin{figure}[htb]%[htb!]
%\hline
%\vspace{0.5cm}
% \centering
%$\lambda=0.9\mu m$\\
 \centering
   \subfigure [\label{histogram_deflection_efficiency_lamb09_teta_80_deflec} $\theta=80^o$, $\lambda=0.9\mu m$]
        {\includegraphics[width=0.22\textwidth]{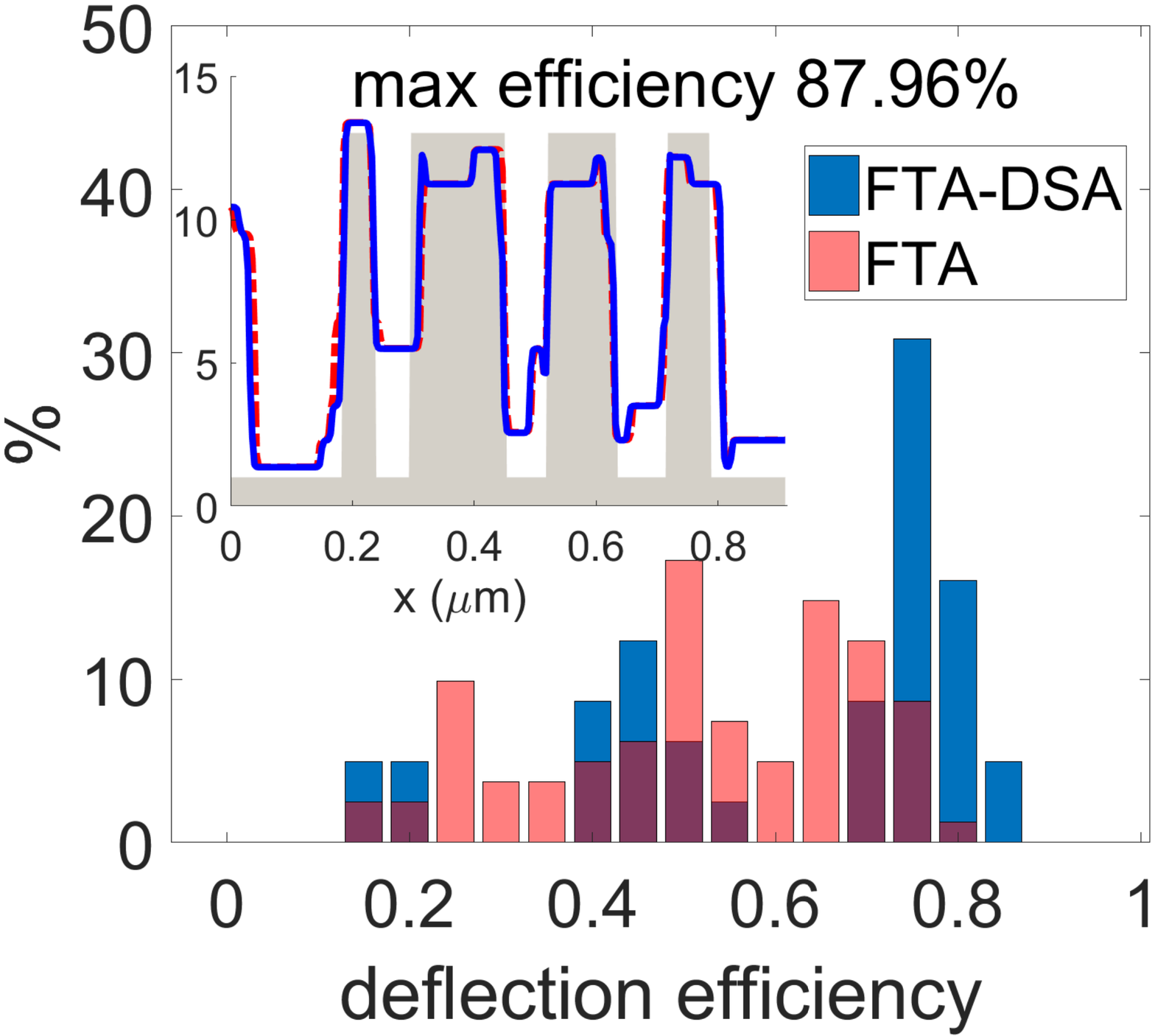}}
        %\caption{\label{deflection_GG_high} toto}
 \centering
 \subfigure [\label{map_histogram_deflection_efficiency_lamb09_teta_80_deflec} $\theta=80^o$, $\lambda=0.9\mu m$]
 {\includegraphics[width=0.22\textwidth]{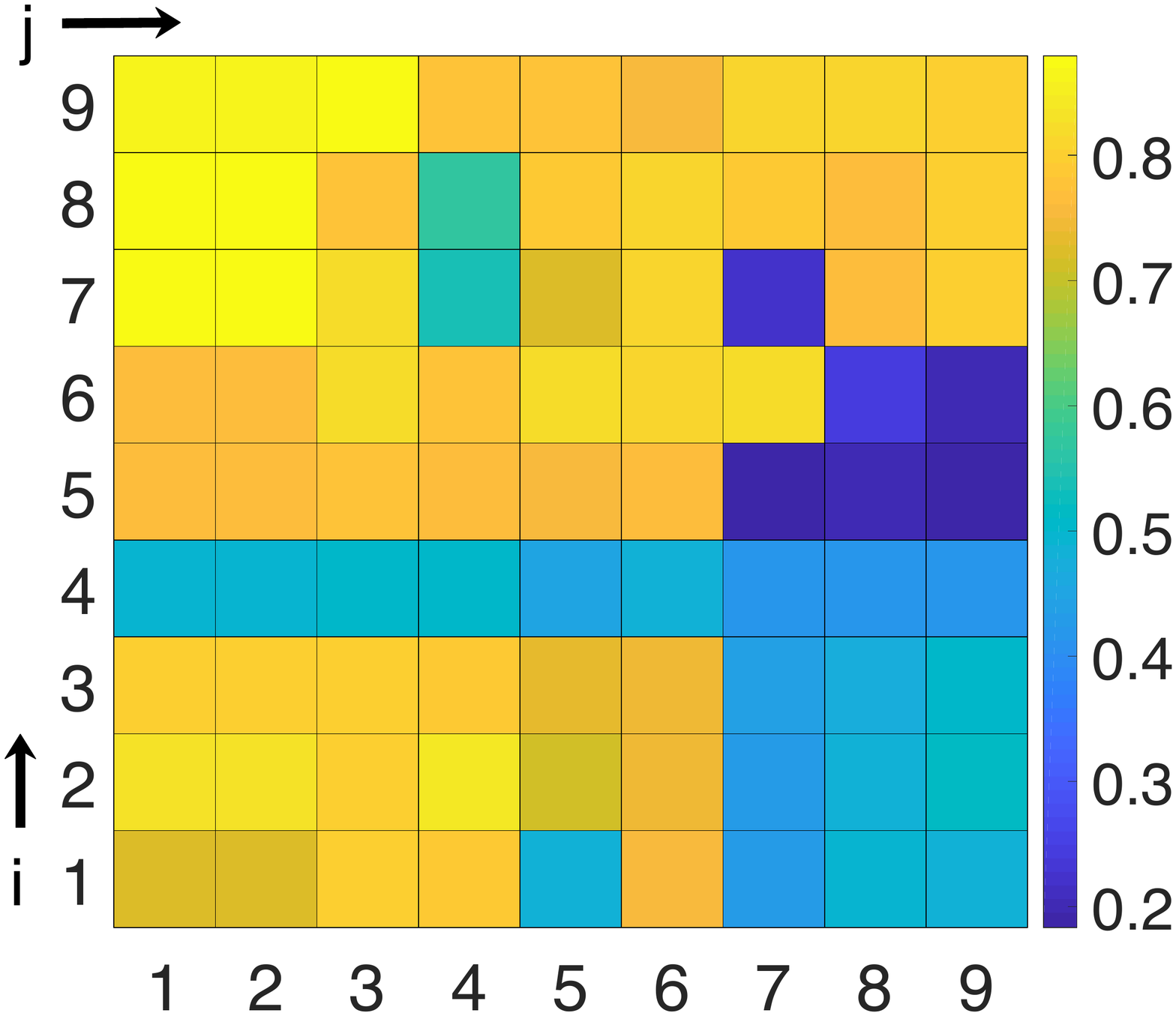}}
%\hline
%\vspace{0.5cm}
%\centering
%$\lambda=1.1\mu m$\\
 \centering
   \subfigure [\label{histogram_deflection_efficiency_lamb11_teta_80_deflec} $\theta=80^o$, $\lambda=1.1\mu m$]
        {\includegraphics[width=0.22\textwidth]{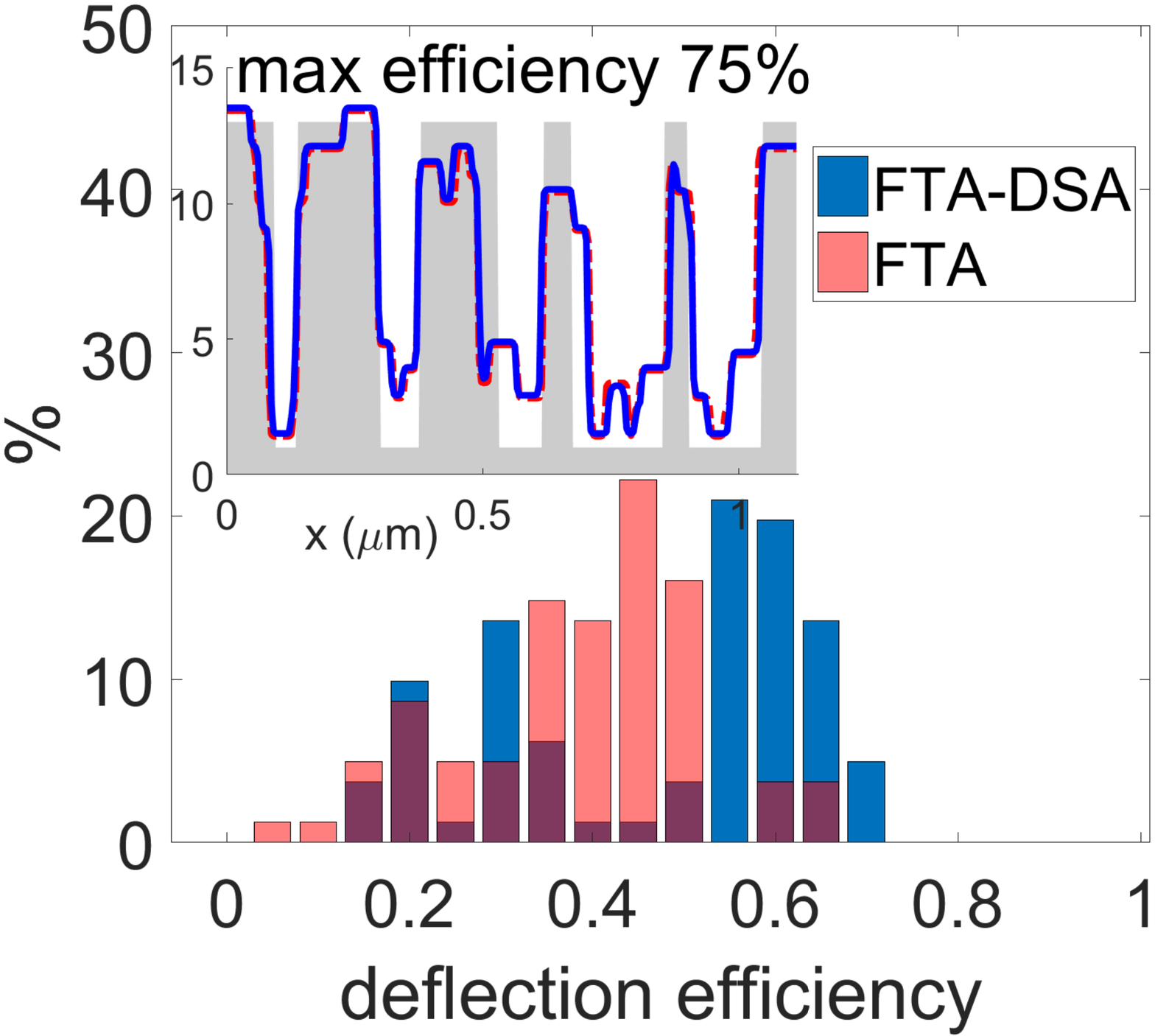}}
        %\caption{\label{deflection_GG_high} toto}
 \centering
 \subfigure [\label{map_histogram_deflection_efficiency_lamb11_teta_80_deflec} $\theta=80^o$, $\lambda=1.1\mu m$]
 {\includegraphics[width=0.22\textwidth]{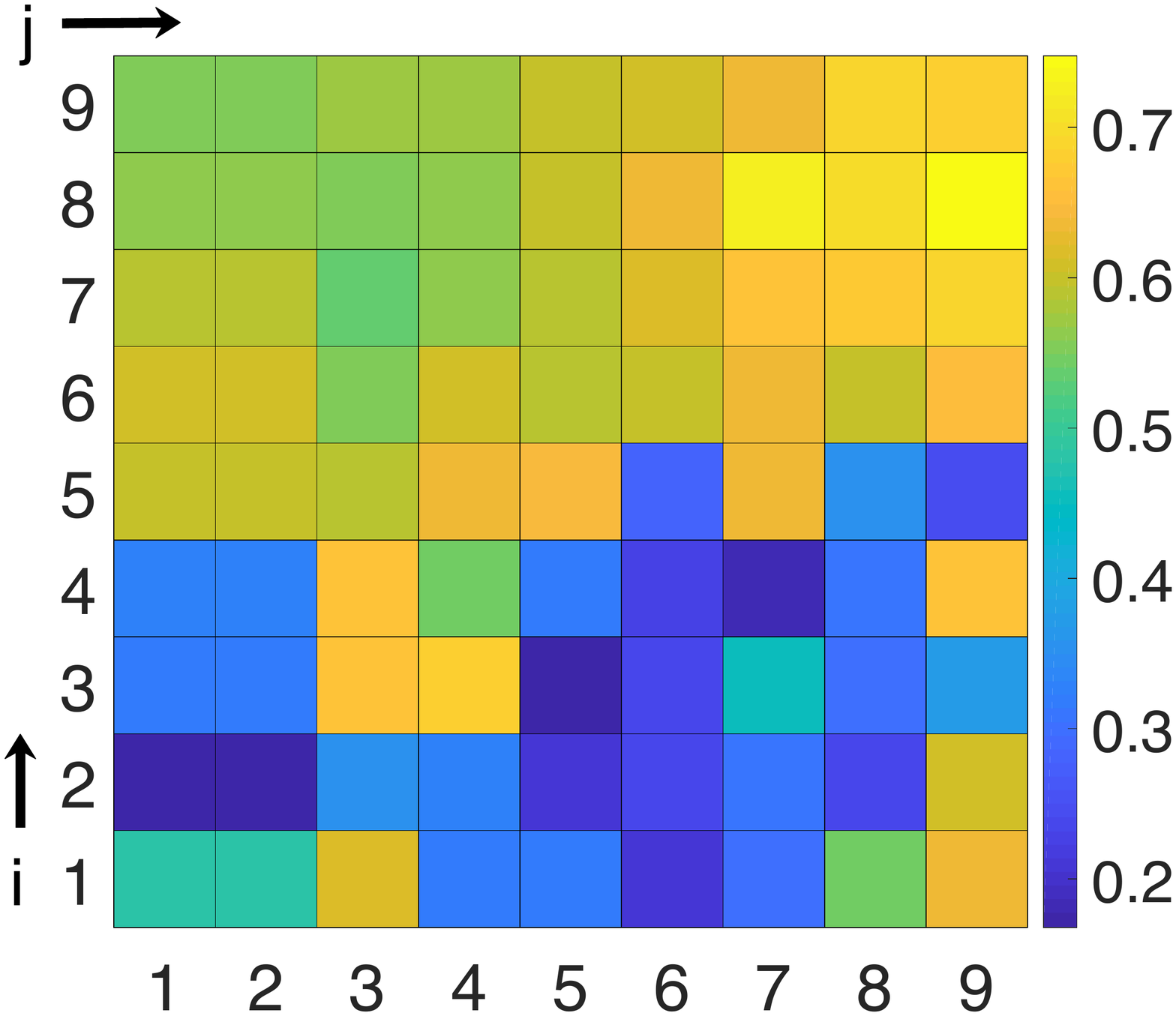}}
\caption{\label{map_histogram_deflection_efficiency_lamb}
TFA-DSA applied to design a 1D high-transmissive metasurface deflecting TM incident planewave onto $\theta_d=80^0$. Figures \ref{histogram_deflection_efficiency_lamb09_teta_80_deflec} and \ref{histogram_deflection_efficiency_lamb11_teta_80_deflec} shows the histogram of optimized results for $\lambda=0.9 \mu m$ and $\lambda=1.1 \mu m$ respectively.  The DSA order is set to $n=4$ yielding a $9 \times 9$ grid. The projection of optimized devices landscape on the DSA 2D array is presented in figure \ref{map_histogram_deflection_efficiency_lamb09_teta_80_deflec} and \ref{map_histogram_deflection_efficiency_lamb11_teta_80_deflec} for $\lambda=0.9\mu m$ and  $\lambda=1.1\mu m$, respectively.}
\end{figure}
%=======================================
\begin{figure}[htb]%[htb!]
 \centering
 \subfigure [\label{histogram_deflection_efficiency_lamb09_teta_neg_60} histogram of efficiencies]
 {\includegraphics[width=0.22\textwidth]{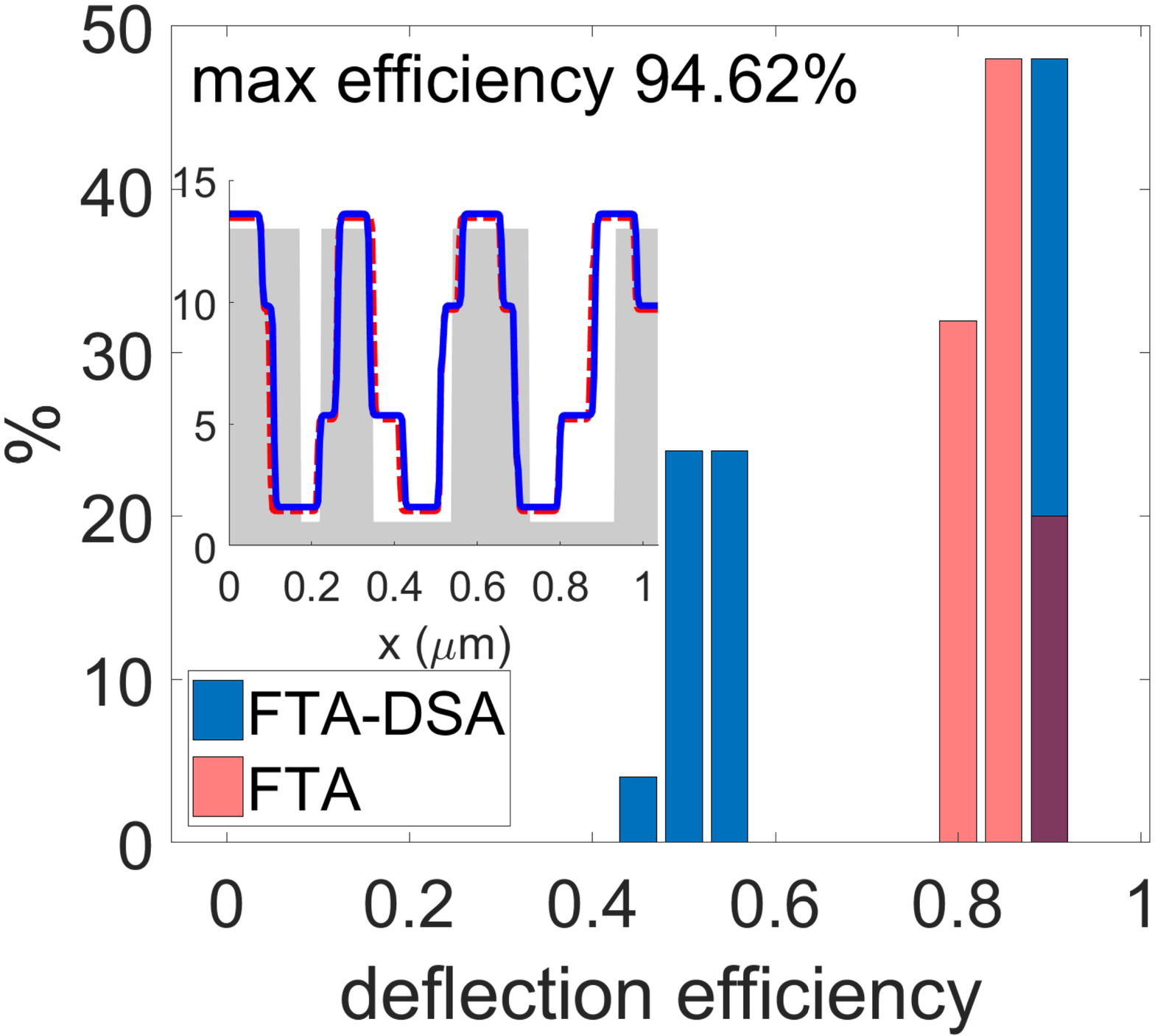}}
 \centering
   \subfigure [\label{filed_map_Hy_neg_refrac} real part of the magnetic field of $\varepsilon^{(5,3)}$ device]
        {\includegraphics[width=0.22\textwidth]{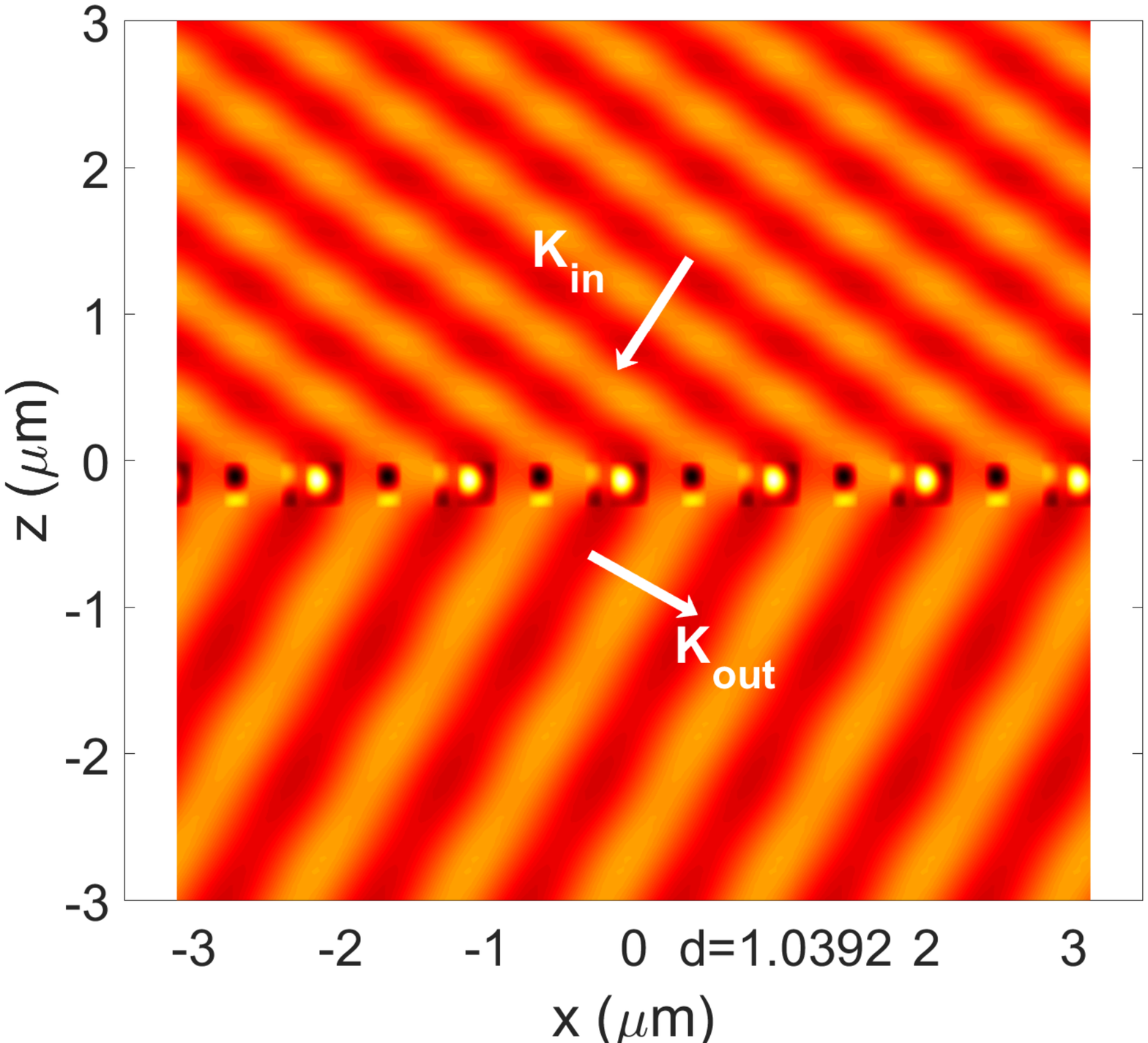}}
%=======================================
\caption{\label{map_histogram_deflection_efficiency_neg}
Illustration of a negative refraction based on the FTA-DSA.  The optimized 1D dielectric metagrating  deflects an incident plane wave with incidence $\theta_{in}=+36.67^o$ into $\theta_{out}=-60^o$. Figure \ref{histogram_deflection_efficiency_lamb09_teta_neg_60} shows the histogram of the optimized  devices and one of the best structure. The father profile is draw in blue and the mother is in dash line red color. Fig. \ref{filed_map_Hy_neg_refrac} shows the real part of the magnetic field in order to illustrate the quality of the  deflection phenomenon supported by one of the final highest-transmission devices.
Numerical parameters: $\lambda=0.9\mu m$, grating period  $d=\nu_3\lambda/sin(\theta_{d})$.}
\end{figure}
The strategy proposed so far consists in  using the DSA only once during the optimization process. The FTA is first used, with a very low number of iterations in order to identify a quartet of  best trends. Then the DNA of this quartet is disseminated through various offspring. Finally, a greater number of iterations of FTA is applied to each member of this new family. This scenario  can be termed one-layer or unsupervised strategy. It seems obvious  to consider a strategy based on a perpetual assessment  of the initial conditions as the algorithm progresses. This strategy  leads to  a multi-layers or a supervised scheme. 
A supervised or multi-layers strategy, can be viewed as a concatenation of several slices/levels of FTA-DSA-FTA.  In each layer, the FTA and DSA algorithms are  performed alternatively, both  with a very low number of iterations, about $5$. First a  FTA with very low numbers of iterations is performed, yielding a selection of the four best mother and father profiles at this iteration. These best profiles are then  used for mapping the four corners of the next new  DSA grid.
A $n$ order  DSA is then applied  leading to a $(2^{n-1}+1)\times (2^{n-1}+1)$  2D  initial individuals array. This new set of individuals are  then used as initial geometries in a next FTA algorithm which is also performed with a low number of iterations. One can expect that this supervised strategy is an optimal scenario allowing  to efficiently select, as earlier as possible, a class of unseen good optimal solutions. We apply the supervised strategy to the same example discussed earlier on table \ref{tab_old_last_mother}. Results are displayed  on table \ref{tab_multi_level}. Recall that for this example, the structure is designed to deflect a TM normally incident plane wave onto $60^o$ deflection angle for $\lambda=0.9\mu m$ and $\lambda=1.1\mu m$. One can remark that, the highest transmissions displayed on both tables are slightly different. For $\lambda=0.9\mu m$, the highest efficiency from the multilevel FTA-DSA is $97.05\%$ while this value reaches $97.42\%$ in the case of classical or one-level FTA-DSA. For $\lambda=1.1\mu m$, the high-performance structure provided by the classical FTA-DSA deflects $78.26\%$ of the incident power while this deflected power reaches $79.37\%$ when the multi-layers strategy is performed. As pointed out earlier, the maximum values of these two tables are hardly different. However, both results landscapes are fairly different. The multi-layers scheme exhibits a higher number of high-performing structures through the optimization process. For $\lambda=0.9 \mu m$ $40\%$ of  devices optimized thanks to the classical FTA-DSA, have a transmission  efficiency greater than $90\%$ while this ratio reaches $56\%$ when the multi-layers scheme is performed. For $\lambda=1.1 \mu m$, $56\%$ of optimized devices based on the classical FTA-DSA has a deflected power higher than $77\%$ while  the FTA-DSA supervised strategy  yields $96\%$.\\
We perform, complementary analysis on devices operating with a more large deflection angle.  Generally, 1D metasurfaces designed to deflect an incident plane wave onto large deflection angles have bad performances.
In this case,  it is difficult to exhibit high-performance structures from a set of random initial conditions since the design space  with a multitude and very close basins of local minima. Consequently,
the gradient-based method faces a multitude of basins of local minima and a huge number of initial candidates is required  before satisfactory solutions are reached. A way to generate initial candidates allowing to broadly and efficiently span the design space consists in increasing the DSA order $n$.   
%Figures $\theta_d=80^o$ and for the two wavelengths: $\lambda=0.9\mu m$ and $\lambda=1.1\mu m$.
Figure \ref{histogram_deflection_efficiency_lamb09_teta_80_deflec} shows the comparison between the  histograms obtained with the FTA and FTA-DSA, of optimized devices for $\lambda=0.9\mu m$ while the histograms corresponding to $\lambda=1.1\mu m$ are reported in fig. \ref{histogram_deflection_efficiency_lamb11_teta_80_deflec}. The highest-transmission device is also displayed on each figure. %Figures \ref{deflection_SDV_80} present results for a deflection angle $\theta_d=80^o$ for $\lambda=0.9 \mu m$ and .$\lambda=1.1 \mu m$. 
The DSA order is set to $n=4$ yielding a $9 \times 9$ grid. The projection of optimized devices landscape on the DSA 2D array is presented in figure \ref{map_histogram_deflection_efficiency_lamb09_teta_80_deflec} and \ref{map_histogram_deflection_efficiency_lamb11_teta_80_deflec} for $\lambda=0.9\mu m$ and  $\lambda=1.1\mu m$, respectively.
It is worth noting, from all these results that the FTA-DSA scheme  enforces the trend towards particular basins of local minima, leading to a partition of devices histogram and  DSA 2D landscape, onto several narrow disjointed sub-bands and sub-sections respectively. Let's also highlight that, for the current deflection angle, namely $\theta=80^o$, the  results obtained from FTA-DSA still  provide systematically better results than the FTA. 
This fact indicates that the algorithm  improves the possibility of the classical FTA to avoid the trapping of undesirable local optimal solutions.\\
Finally, in order to justify the generalization of the proposed approach, staying in the case of 1D functional dielectric metagrating optimization,  we apply the concept to successfully realize a ultra-high-efficiency anomalous refraction with a 1D dielectric metagrating \cite{Sell2}. As highlighted by Snell et al. in \cite{Sell2}, the realization of metagratings handling this physical phenomenon for arbitrary input (incident) and output angles remains a challenge. Here, we numerically show that the proposed optimization method allows 
to efficiently design ultra-high-efficiency anomalous refraction dielectric metagratings and can yield  arbitrary input and output angles. We consider the same structure used in the previous study and it is designed to deflect $\theta_{in}=+36.67^o$ TM incident planewave to $\theta_{out}=-60^o$ output planewave at $\lambda=0.9\mu m$ operating wavelength. The configuration of the study is displayed in fig. \ref{sketch_FTA_DSA_neg_refraction}. 
Figure \ref{histogram_deflection_efficiency_lamb09_teta_neg_60} shows the histogram of the  devices optimized thanks to the FTA-DSA. FTA-DSA is performed on a $5\times 5$ DSA grid. Nearly $50\%$ of the optimized structures transmit more than $90\%$ of the incident energy.
The optimal structures are a 3-nanorods type with a deflected efficiency close to  $95\%$. The quality of the  deflection phenomenon supported by one of the final highest-transmission devices is illustrated in fig. \ref{filed_map_Hy_neg_refrac} where we  plot the real part of the magnetic field. The incident and deflected  wavefronts  are well-distinguished.
\section{Conclusion}
We design a functional 1D dielectric metasurface based on a topology   optimization method.  The proposed algorithm is based on the fluctuation and trends analysis which initially performs, randomly, a very large class of non-intuitive solutions in the design space. The trend is a simple shape deterministic oscillatory function while a set of  optimal  fluctuating initial geometries are generated through a random Gaussian process with suited  bandlimited correlation function spectrum. 
A suited choice of the trend feature allows to efficiently target optimal solutions. Since, no  method can efficiently predict the trend  profile features, we suggest a randomization of the number of oscillation of the trend function. 
A randomness exploration  process begins by tracking the best initial candidates through a low number of FTA-iterations.  
Using a diamond-square-algorithm (DSA), the best initial candidates family is extended to higher quality offspring that handle the DNA feedback.  
We successfully apply the method to design a 1D metagrating that deflects an incident TM polarized wave into several angles for different wavelengths.  We showed that, the probability to reach a high-performance structure is highly increased, despite the full randomization of the initial profile mean features. 

\section*{Funding Information}
This work has been sponsored by the French government research program "Investissements d'Avenir" through the IDEX-ISITE initiative 16-IDEX-0001 (CAP 20-25)
\bigskip

\section*{Disclosures}
The authors declare no conflicts of interest.

%%%%%%%%%%%%%%%%%%%%%%% References %%%%%%%%%%%%%%%%%%%%%%%%%

\end{document}